\documentclass[twocolumn,trackchanges]{aastex631}
\usepackage{threeparttable}
\usepackage{array,multirow,makecell}
\usepackage{rotating,tabularx}
\usepackage{tablefootnote}
\usepackage{appendix}
\usepackage{rotating}
\usepackage{afterpage}
\newcommand{\msol}{\,M$_\odot$} %solar mass

\submitjournal{ApJ}

\shortauthors{Bouvier et al.}
\shorttitle{ORANGES II. Are ORANGES different from PEACHES?}

\graphicspath{{./}{figures/}}

\begin{document}

\title{The chemical nature of Orion protostars: Are ORANGES different from PEACHES? ORANGES II.} %\footnote{Released on March, 1st, 2021}}

\author[0000-0003-0167-0746]{Mathilde Bouvier}
\affiliation{Univ. Grenoble Alpes, CNRS, IPAG, 38000 Grenoble, France\\}
\email{mathilde.bouvier@univ-grenoble-alpes.fr}

\author[0000-0001-9664-6292]{Cecilia Ceccarelli}
\affiliation{Univ. Grenoble Alpes, CNRS, IPAG, 38000 Grenoble, France\\}

\author[0000-0002-6729-3640]{Ana L\'opez-Sepulcre}
\affiliation{Univ. Grenoble Alpes, CNRS, IPAG, 38000 Grenoble, France\\}
\affiliation{Institut de Radioastronomie Millimétrique (IRAM), 300 rue de la Piscine, F-38400 Saint-Martin d’Hères, France}

\author[0000-0002-3297-4497]{Nami Sakai}
\affiliation{RIKEN Cluster for Pioneering Research, 2-1 Hirosawa, Wako-shi, Saitama 351-0198, Japan}

\author[0000-0002-9865-0970]{Satoshi Yamamoto}
\affiliation{Department of Physics, The University of Tokyo, 7-3-1, Hongo, Bunkyo-ku, Tokyo 113-0033, Japan}
\affiliation{Research Center for the Early Universe, The University of Tokyo, 7-3-1, Hongo, Bunkyo-ku,
Tokyo 113-0033, Japan}

\author[0000-0001-8227-2816]{Yao-Lun Yang}
\affiliation{RIKEN Cluster for Pioneering Research, 2-1 Hirosawa, Wako-shi, Saitama 351-0198, Japan}
\affiliation{Department of Astronomy, University of Virginia, Charlottesville, VA 22904-4235, USA}

\begin{abstract}

Understanding the chemical past of our Sun and how life appeared on Earth is no mean feat. The best strategy we can adopt is to study newborn stars located in an environment similar to the one in which our Sun was born and assess their chemical content. In particular, hot corinos are prime targets since recent studies showed correlations between interstellar Complex Organic Molecules (iCOMs) abundances from hot corinos and comets. The ORion ALMA New GEneration Survey (ORANGES) aims to assess the number of hot corinos in the closest and best analogue to our Sun's birth environment, the OMC-2/3 filament. In this context, we investigated the chemical nature of 19 solar-mass protostars and found that 26\% of our sample sources shows warm methanol emission indicative of hot corinos. Compared to the Perseus low-mass star-forming region, where the PErseus ALMA CHEmistry Survey (PEACHES) detected $\sim 60$\% of hot corinos, the latter seem to be relatively scarce in the OMC-2/3 filament. While this suggests that the chemical nature of protostars in Orion and Perseus is different, improved statistics is needed in order to consolidate this result. If the two regions are truly different, this would indicate that the environment is likely playing a role in shaping the chemical composition of protostars. 

\end{abstract}


\section{Introduction} \label{sec:intro}

Understanding how life appeared on Earth is one of the Holy Grail in Science. From the astrophysical point of view, the Sun's birth environment being long dissipated, we cannot see what happened in its youth. We can, however, study solar-mass protostars that are currently forming in other regions of our Galaxy to understand the full story of our planetary system formation. 

The discovery of two chemically distinct types of solar-mass protostars, hot corinos and Warm Carbon Chain Chemistry (WCCC) sources, shows that the story might not be the same for every solar-mass protostar. While hot corinos are compact ($\leq$100 au), hot ($\geq 100$ K), and dense ($\geq 10^{7}$cm$^{-3}$) regions \citep{ceccarelli2004, ceccarelli2007}, enriched in interstellar Complex Organic Molecules (iCOMs; \citealt{herbst2009, ceccarelli2017}), WCCC objects are deficient in iCOMs but show a larger zone ($\sim$ 2000 au) enriched in unsaturated carbon chain molecules \citep{sakai2008b,sakai2013}. In between these two extreme cases, there exist objects called hybrids that present both hot corino and WCCC features (e.g. L483, B335; \citealt{imai2016, oya2017, jacobsen2019}).

Until recently, only a dozen hot corinos were discovered, but thanks to the arrival of powerful (sub)-mm interferometers such as ALMA, more hot corinos are identified. In particular, the recent Perseus ALMA Chemistry Survey (PEACHES; \citealt{yang2021}) targeted 50 solar-mass protostars in the Perseus Molecular Cloud, a region forming only low-mass stars. They found that $\sim$56\% of their source sample show warm methanol emission, indicating that hot corinos are likely prevailing in this region. The Perseus Molecular Cloud is, however, different from the Solar birth environment. The latter was most likely a dense protocluster with high-mass stars in its vicinity \citep[e.g.][]{adams2010, pfalzner2015}. Are hot corinos also abundant in an environment analogue to that where our Sun was born? Recent studies showed similarities between the abundances of iCOMs found in hot corinos compared to those found in comets \citep[][]{bianchi2019,drozdovskaya2019,rivilla2020}. Did our Sun experience a hot corino phase? We need to target low-mass protostars belonging to massive star-forming regions (SFRs).  

The closest and best analogue of our Sun's birth environment is the OMC-2/3 filament, located in the Orion A molecular cloud. Very recently, three hot corinos were detected in this region, the intermediate-mass protostars HOPS-87 (also known as MMS6) and HOPS-370 (also known as OMC2-FIR3) \citep{hsu2020, tobin2019} and the solar-type protostar HOPS-108 located in the OMC-2 FIR4 protocluster \citep{tobin2019, chahine2022}. Although hot corinos are present in massive SFRs \citep{codella2016, hsu2020, chahine2022}, the statistics is too poor to draw any conclusion on the chemical past of our Sun. We, therefore, need more systematic studies of hot corinos in massive SFRs.

The ORion ALMA New GEneration Survey (ORANGES) is a project aiming to study the chemical nature of the Solar-type protostars located in the OMC-2/3 filament, (393 $\pm$ 25) pc from the Sun \citep{grosschedl2018}, with an angular resolution of 0.25$''$ ($\sim $100 au). ORANGES is analogous to PEACHES because the two studies have been designed to have the same sensitivity (corrected for the distance), spatial resolution and spectral setup. It allows a direct comparison of the two environments, i.e. the OMC-2/3 filament and the Perseus Molecular Cloud. One of the goals of ORANGES is to assess the number of hot corinos in the OMC-2/3 region, and provide a first answer concerning the chemical past of our Sun. In ORANGES, we targeted the same protostars targeted by \citet{bouvier2021}. They were initially 9 chosen protostellar sources based on single-dish studies \citep[e.g.][]{chini1997,lis1998,nielbock2003} satisfying the following three criteria: (1) detection in the (sub-)mm continuum emission; (2) estimated envelope mass $\leq 12$ \msol; (3) bona fide Class 0 and I protostars (see \citealt{bouvier2020}). The recent interferometric studies showed that most of these systems are in fact multiple systems \citep[][]{tobin2020, bouvier2021} which led to a total number of 19 studied targets.

The results of a previous single-dish study \citep[][]{bouvier2020} towards the same targets showed that the large scale ($\leq 10^4$ au) line emission is dominated by the Photo-Dissociation Region (PDR) or by the molecular cloud, rather than the protostellar envelopes. Interferometric observations are thus essential to detect hot corinos in this highly illuminated region. In this study, we investigated the most common tracer of hot corinos, CH$_3$OH, in a sample of 19 embedded solar-type protostars. Table \ref{tab:sources-sample} lists the targeted protostars and their coordinates.

\begin{figure*}[ht]
    \centering
    \includegraphics[width=0.9\linewidth]{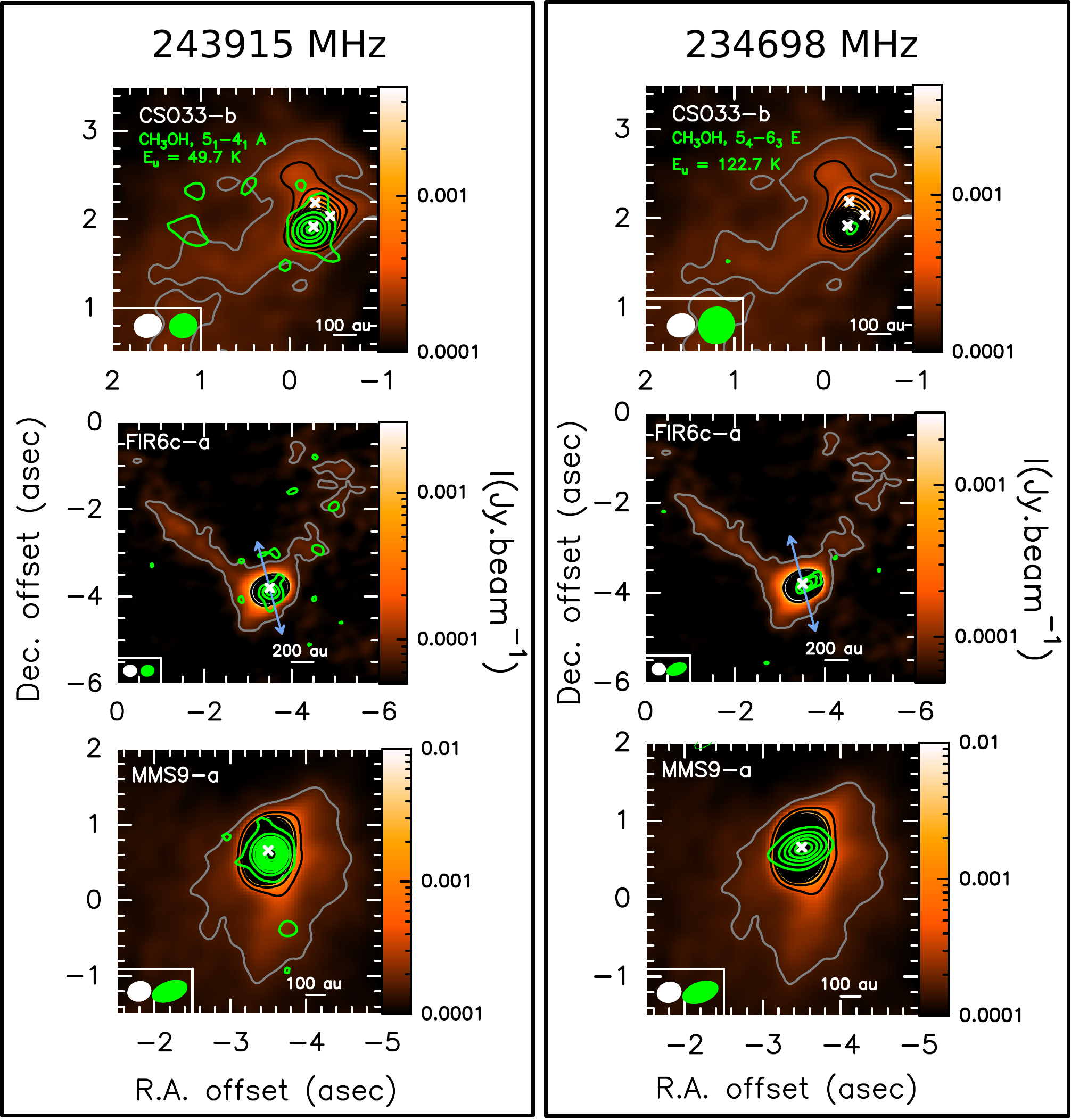}
    \caption{1.3mm continuum maps of CSO33-b-a, FIR6c-a, and MMS9-a (coloured area, grey and black contours). The first contour levels are in grey. Levels start at 15$\sigma$ for CSO33-b-a (1$\sigma=44\mu$Jy/beam) and FIR6c-a (1$\sigma=60\mu$Jy/beam), and 20$\sigma$ for MMS9-a (1$\sigma=50\mu$Jy/beam). Level steps are 50$\sigma$ except for CSO33-b-a where the step is 10$\sigma$. The moment 0 emission of the CH$_3$OH transitions at 243915 MHz ($E_u=$ 49.7 K) and at 234698 MHz ($E_u=$ 122.7 K) are shown with green contours in the left and right columns, respectively. For the 243915 MHz transition line, contours start at 3$\sigma$ (1$\sigma= 6, 9, 8$ mJy.beam$^{-1}$.km.s$^{-1}$ for CSO33-b-a, FIR6c-a, and MMS9-a, respectively) with steps of 3$\sigma$ for CSO33-b-a and FIR6c-a, and steps of 5$\sigma$ for MMS9-a. For the 234698 MHz transition line, contours start at 3$\sigma$ (1$\sigma= 9, 9, 12$ mJy.beam$^{-1}$.km.s$^{-1}$ for CSO33-b-a, FIR6c-a, and MMS9-a, respectively) with steps of 1$\sigma$ for CSO33-b-a, FIR6c-a and  $3\sigma$ for MMS9-a. The continuum and methanol associated synthesised beams are in white and green respectively and are depicted in the lower left corner of the boxes. Light blue arrows represent the orientation of the outflow of the source when known \citep[e.g.][]{williams2003, takahashi2008,shimajiri2009, tanabe2019,gomezruiz2019, feddersen2020}. White crosses represent the position of the sources. }
    \label{fig:cont_ch3oh_part1}
\end{figure*}

\begin{figure*}[ht]
    \centering
    \includegraphics[width=0.9\linewidth]{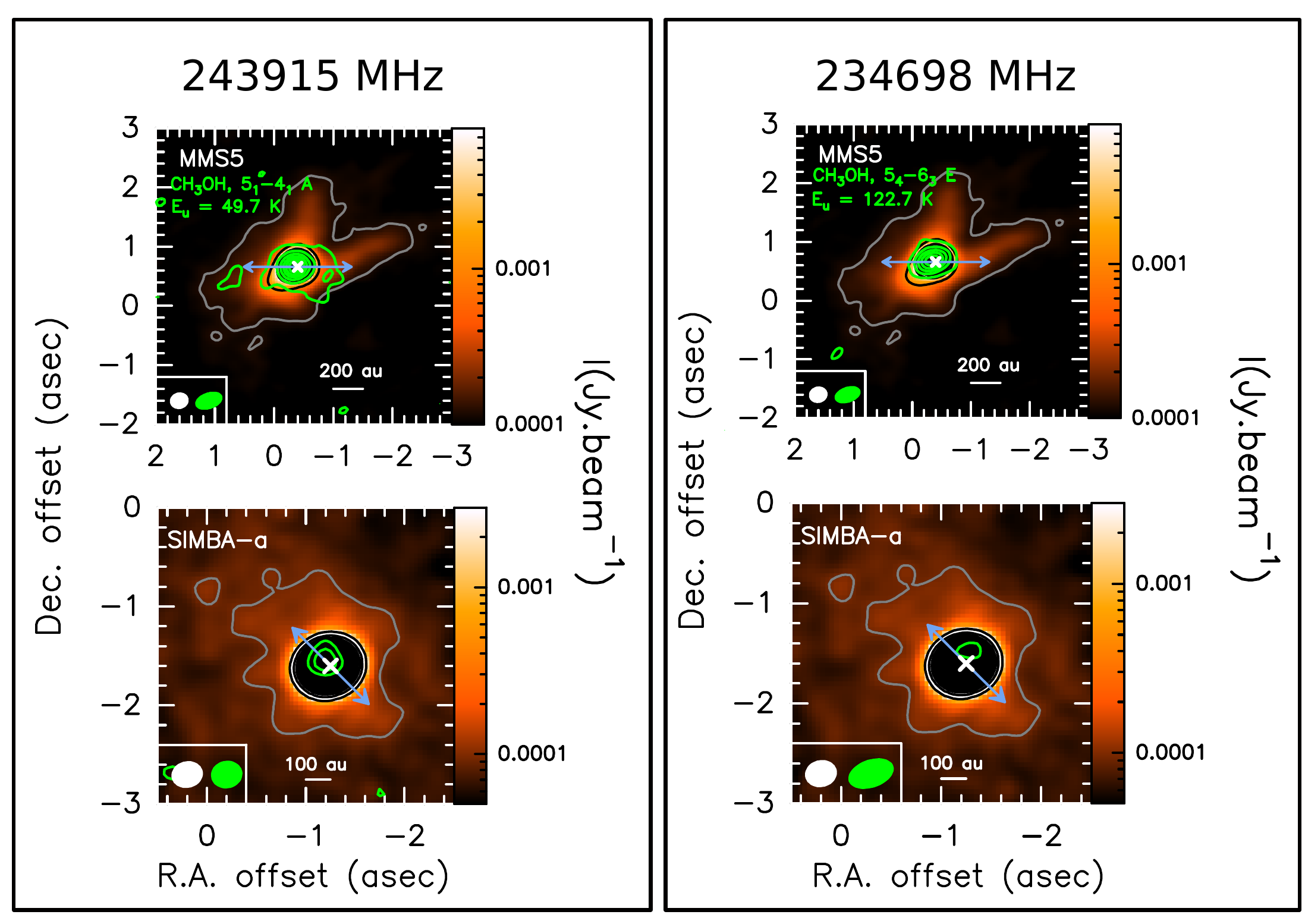}
    \caption{1.3mm continuum maps for MMS5 and SIMBa-a (coloured area, grey and black contours). The first contour levels are in grey. Levels start at 15$\sigma$ for MMS5 (1$\sigma=80\mu$Jy/beam) and 10$\sigma$ for SIMBA-a (1$\sigma=50\mu$Jy/beam). Level steps are 50$\sigma$. The moment 0 emission of the CH$_3$OH transitions at 243915 MHz ($E_u=$ 49.7 K) and at 234698 MHz ($E_u=$ 122.7 K) are shown with green contours in the left and right columns, respectively. For the 243915 MHz transition line, contours start at 3$\sigma$ (1$\sigma=$ 7 and 5 mJy.beam$^{-1}$.km.s$^{-1}$ for MMS5 and SIMBA-a, respectively) with steps of 10$\sigma$ for MMS5 and 1$\sigma$ for SIMBA-a. For the 234698 MHz transition line, contours start at 3$\sigma$ (1$\sigma=$ 10 and 7 mJy.beam$^{-1}$.km.s$^{-1}$ for MMS5 and SIMBA-a, respectively) with steps of $5\sigma$ for MMS5 and 1$\sigma$ for SIMBA-a. The continuum and methanol associated synthesised beams are in white and green respectively and are depicted in the lower left corner of the boxes. Light blue arrows represent the orientation of the outflow of the source when known \citep[e.g.][]{williams2003, takahashi2008, tanabe2019,gomezruiz2019,matsushita2019, feddersen2020}. White crosses represent the position of the sources. }
    \label{fig:cont_ch3oh_part2}
\end{figure*}

\section{Observations} \label{sec:obs}

The observations were performed between 2016 October 25th and 2017 May 5th during Cycle 4, under the ALMA project 2016.1.00376.S. The observations were performed in Band 6 using two different spectral setups. The ranges of frequencies covering the methanol transitions relevant for this work are 243.88 -- 243.97 GHz and 261.77 -- 261.88 GHz for setup 1, and 218.38 -- 218.50 GHz, 230.33 -- 234.08 GHz, and 234.64 -- 234.76 GHz for setup 2. For setup 1, a total of 41 antennas of the 12-m array were used with a baseline length range of 18.6m -- 1100m. The  integration time is $\sim$ 20 min per source. For setup 2, a total of 45 antennas of the 12-m array were used with a baseline range of 18.6m -- 1400m. The integration time is $\sim 8$ min per source. The ALMA correlator was configured to have both narrow and wide spectral windows (spws), with 480 and 1920 channels respectively. Narrow spws have a bandwidth of 58.59 MHz with a channel spacing of 122 kHz ($\sim$ 0.15-0.17 km/s) while the wide spws have a bandwidth of 1875 MHz with a channel spacing of 0.977 MHz ($\sim$ 1.2-1.3 km/s). The bandpass and flux calibrators were J0510+1800 and J0522-3627, and the phase calibrators were J0607-0834 and J0501-0159. The flux calibration error is estimated to be better than 10\%. The precipitable water vapour (PWV) was typically less than 1mm and the phase root-mean-square (rms) noise less than 60$^{\circ}$. In the context of the ORANGES project, several molecular species were targeted but we focus here in particular on methanol (CH$_3$OH), the typical tracer of hot corinos. The methanol lines were found in six (both narrow and wide) spws. The rest frequencies of the methanol transition lines and the associated primary beam sizes are shown in Table \ref{tab:methanol-spw}. \\

We used the Common Astronomy Software Application (CASA; \citealt{mccmullin2007}) for the data calibration. We then exported the calibrated visibility tables to GILDAS \footnote{\url{http://www.iram.fr/IRAMFR/GILDAS}} format and performed the imaging in MAPPING. We first produced a continuum image by averaging line-free channels in the visibility plane using an automatic procedure. We then subtracted the continuum from the line emission directly in the visibility plane. We cleaned the cubes using natural weighting (with the CLEAN procedure) down to $\sim 24$mJy/beam on average. The phase self-calibration performed on the continuum of the sources \citep[see][]{bouvier2021} has been applied to the cubes. The narrow spws were re-sampled to a channel spacing of 0.5 km.s$^{-1}$. The maps shown in this paper are not corrected for the primary beam attenuation but we took into account the correction to measure the line intensities. The resulting synthesized beam and rms for each source and each spectral window are presented in Table \ref{tab:spectral-params}.

\section{Results} \label{sec:results}

\subsection{Methanol Lines}\label{subsec:methanol}

\begin{figure*}[ht]
    \centering
    \includegraphics[width=0.8\linewidth]{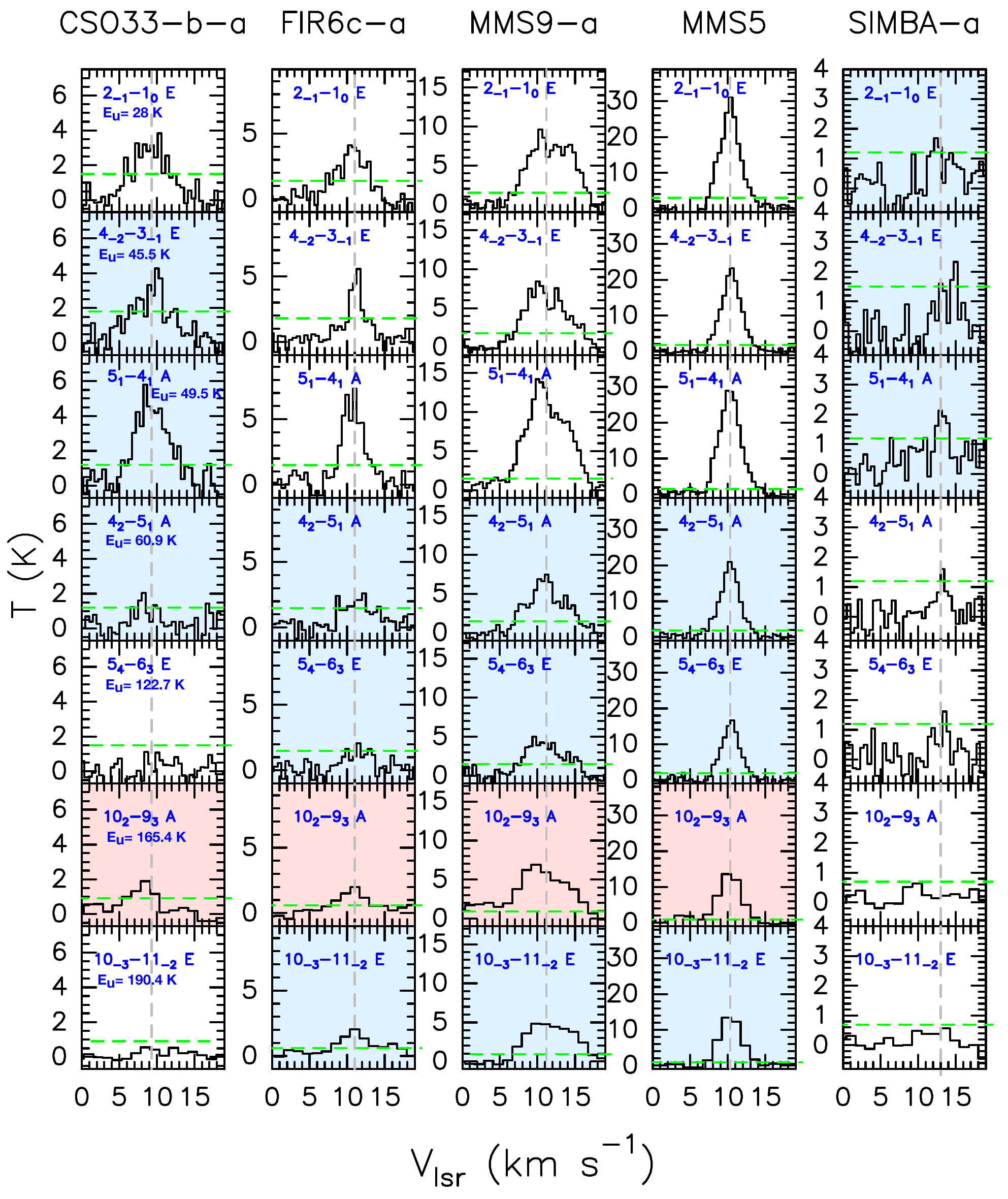}
    \caption{Methanol spectral lines detected in each source. The lines taken into account in the LVG analysis are with the blue background. The lines with the red background are likely contaminated by a line of $^{33}$SO$_2$ and are thus left out from the LVG analysis. The transition of each line is marked in the top left corner of the boxes. Dashed green lines show the 3$\sigma$ level and dashed grey lines the averaged fitted peak velocity of all transitions of the associated source, $V_{\rm peak}$, determined from the Gaussian line fitting.}
    \label{fig:spectra}
\end{figure*}

Methanol is detected towards the centre of 5 out of the 19 protostars: CSO33-b-a, FIR6c-a, MMS9-a, MMS5 and SIMBA-a. In these sources, the line spectra were extracted from the pixel corresponding to the position of the methanol peak, which often corresponds to the continuum emission peak. The coordinates of the position where the spectra have been extracted are indicated in Table~\ref{tab:sources-sample}. The line detection threshold is set to $3\sigma$ at the line emission peak. Figures \ref{fig:cont_ch3oh_part1} and \ref{fig:cont_ch3oh_part2} show the moment 0 map of the two CH$_3$OH lines at 243915 MHz and 234698 MHz, which have different upper-level energies $E_u$, overlaid on the 1.3mm dust continuum emission of each source. We note that for CSO33-b-a and SIMBA-a, the methanol transition at 234698 MHz ($E_u=$ 122.7 K) is considered as undetected as the emission is shown only by a 3$\sigma$ contour which is not centred on the source's continuum peak. We found that whilst the emission of methanol lines with low upper-level energy, such as the 243915 MHz transition, is resolved and extended in most sources, the emission of methanol lines with high upper-level energy, such as the 234698 MHz transition, is compact. Methanol emission is seen near other sources of the sample but not at the position of the protostars. As we are interested in detecting hot corinos, we will focus in this letter only on the 5 sources cited above.

We detected up to eleven CH$_3$OH lines with upper level energies $E_{\text{up}}$ from 28 to 537 K and Einstein coefficients $A_{ij}$ between $6.3 \times 10^{-6}$ and $1 \times 10^{-4}$ s$^{-1}$. The extracted spectra of methanol lines for each source are shown in Fig. \ref{fig:spectra}. We performed a Gaussian line fitting to each source in order to extract the line width (FWHM) and the peak velocity ($V_{\rm peak}$). To extract the integrated intensity, we did a Gaussian fit ($\int T_{B}dV$ G.) and we also measured it by direct integration of the channel intensities ($\int T_{B}dV$ D.). Only MMS5 has lines with Gaussian profiles so we used the Gaussian fit results for this source and the results of the direct integration for the other sources. The line fitting results are reported in Table \ref{tab:spectral-params}, as well as the rms computed for each spectral window. Line widths range between $\sim$ 2 and 7 km.s$^{-1}$.

Methanol lines can be very optically thick towards hot corinos \citep{bianchi2020}. We therefore looked for the isotopologue CH$_3^{18}$OH which is usually optically thin, in order to derive the methanol column density more accurately. Among the seven CH$_3^{18}$OH lines expected to be the most intense, we detected and used only one line. The other lines are either undetected ($\leq 3\sigma$), or contaminated by lines from other molecules such as C$_2$H$_5$OH, C$_2$H$_5$CN, or CH$_2$DOH. The spectral parameters and Gaussian fit results of the transition used in this work, which is the $5_{0,5} - 4_{0,4}$ A transition at 231758 MHz, are reported in Table \ref{tab:spectral-params}. The frequencies of the seven CH$_3^{18}$OH spectral lines expected to be the most intense are indicated in Fig. \ref{fig:ch3oh18h}.

\subsection{Non-LTE LVG Analysis}\label{subsec:LVG}
To derive the physical properties of the gas where methanol is emitted, we performed a non-LTE analysis using the Large Velocity Gradient(LVG) code \textit{grelvg}, originally developed by \citet{ceccarelli2003}. We used the CH$_3$OH-H$_2$ collisional rates from \citet{flower2010} between 10 and 200 K for the first 256 levels, provided by the BASECOL database\footnote{\url{https://basecol.vamdc.eu/}} \citep{dubernet2013}. We assumed a spherical geometry to compute the line escape probability \citep{dejong1980}, a ratio CH$_3$OH-E/CH$_3$OH-A equal to 1, and an H$_2$ ortho-to-para ratio of 3. The assumed line widths are those measured from the spectral lines towards each source (see Table \ref{tab:spectral-params}) and we included the calibration error of 10\% in the observed intensities. 

The detected methanol transitions span a large range of $E_{\text{up}}$. First, methanol lines with $E_{\text{up}}$ higher than 400 K have been excluded from the analysis as the collisional coefficients are not computed at these energies. Second, low energy transitions can eventually trace a different region than the higher energy level transitions. Indeed, the low upper energy level transitions ($E_u \leq 50$ K) show extended emission towards most of the sources, while the high upper level ones are compact. We, therefore, did not consider the low upper energy lines when performing the LVG analysis, except for CSO33-b-a and SIMBA-a where we detected only three low level energy transitions. Additionally, the line at 232418 MHz ($E_u=$ 165 K) is likely contaminated by a $^{33}$SO$_2$ line falling at the same frequency. We do not have enough information (i.e. other lines) to evaluate the possible contribution of this line. We, thus, excluded this line from the LVG analysis as well.

In the case of MMS5, we also included the detected line of CH$_3^{18}$OH-A with the $^{18}$O/$^{16}$O ratio equal to 560 \citep[][]{WR94} to better constrain the derived total CH$_3$OH column density for this source. For each source, the lines that are \textit{not} used for the LVG analysis are shown in \textit{italic} in Table~\ref{tab:spectral-params}. In most cases, we ran the LVG radiative transfer code with only three lines so  that the accuracy of the fit is not very elevated.

For each source we ran a large grid of models varying the total (CH$_3$OH-E + CH$_3$OH-A) column density from $2 \times 10^{14}$ cm$^{-2}$ to $3 \times 10^{19}$ cm$^{-2}$, the gas temperature from 20 to 200 K, and the H$_2$ density from $3 \times 10^{5}$ cm$^{-3}$ to $1 \times 10^{10}$ cm$^{-3}$. These ranges for the parameters are those expected in hot corinos and in outflows shocks, as we expect the emission coming from either of these two types of environments. We fitted the measured CH$_3$OH-E and CH$_3$OH-A lines intensities simultaneously via comparison with the LVG model predictions, leaving $N_{\text{CH3OH}}$, $n_{\text{H2}}$, $T_{\text{kin}}$ and the source size ($\theta$) as free parameters. Then, since the lines are optically thin in the cases of CSO33-b-a and SIMBA-a, there is a degeneracy between the source size and the column density and the best fit of the LVG analysis actually provides the product $\theta \times N_x$. For these sources, we re-ran the best-fitting procedure, this time by fixing the source size and leaving $N_{\text{CH3OH}}$, $n_{\text{H2}}$, and $T_{\text{kin}}$ as free parameters. We then varied the source size around its best-fit value to find when the $\theta \times N_x$ product does not give the same chi square, namely, where the degeneracy disappears.

\begin{figure}
    \centering
    \includegraphics[width=1\linewidth]{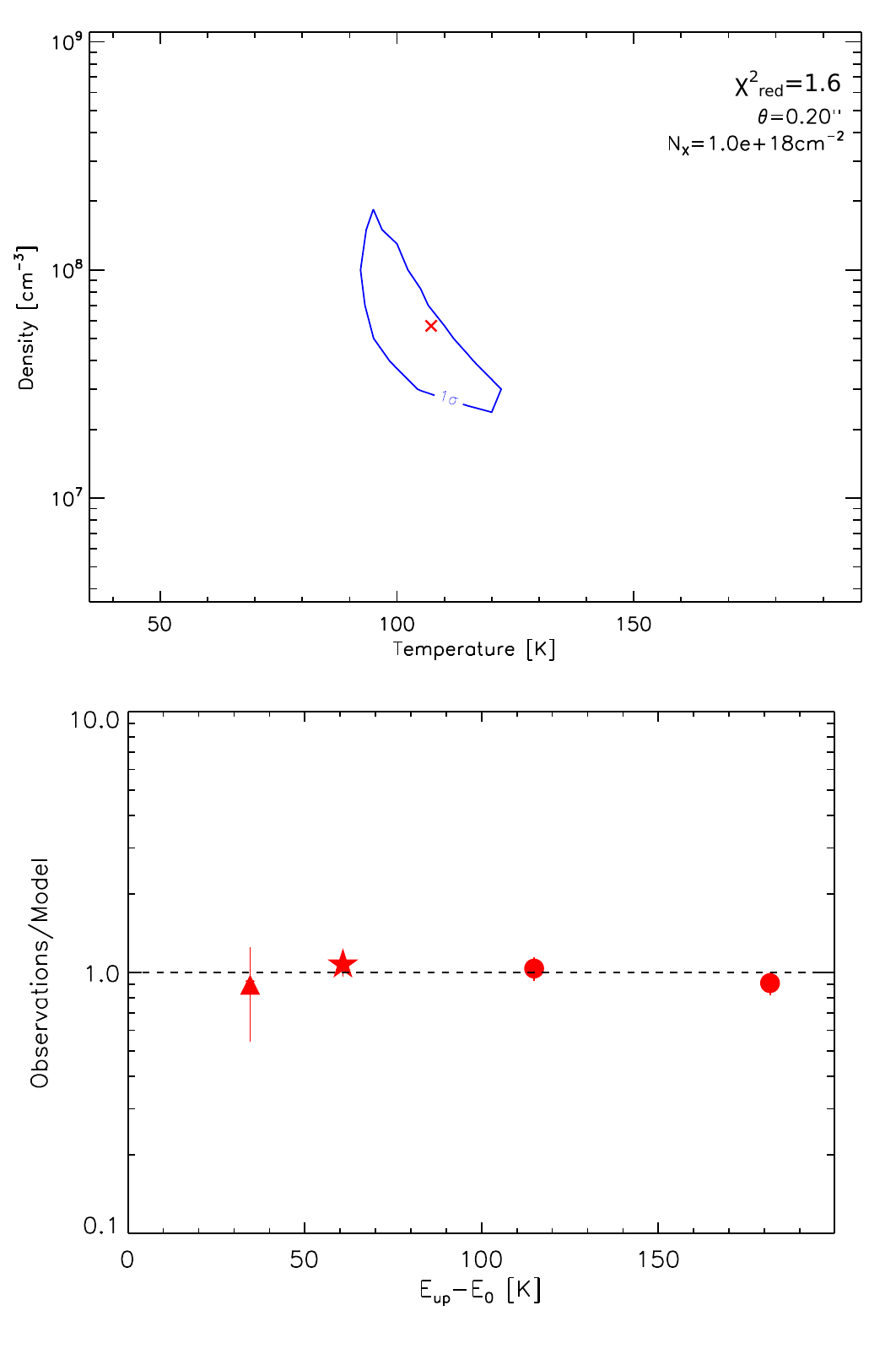}
    \caption{Result of the LVG for MMS5. \textit{Top:} Density-temperature $\chi^2$ contour plot. The best-fit solution is marked by a red star and the blue contours represent the 1$\sigma$ confidence level, assuming the best-fit values for N$_{\text{CH3OH-E}}$ and $\theta$ (Table \ref{tab:results_LVG}). \textit{Bottom:} Ratio between the observed line intensities with those predicted by the best fit model as a function of the line upper level energy $E_{\text{u}}$. Circles and stars refers to CH$_3$OH-E and CH$_3$OH-A respectively, whilst the triangle refers to the CH$_3^{18}$OH-A detected line. E$_0$ is the ground rotational level energy which is 0 K for CH$_3$OH-A and 7.7 K for CH$_3$OH-E \citep[][]{flower2010}. }
    \label{fig:LVG_res}
\end{figure}

The best fit for the total CH$_3$OH column densities range between 8$\times 10^{15}$ and 4$\times 10^{18}$ cm$^{-2}$ with reduced $\chi^2_{red}$ between 0.1 and 1.6. All the lines for CSO33-b-a, SIMBA-a, and the CH$_3^{18}$OH line for MMS5 are optically thin ($\tau_L \leq 1$; $\tau_L$ being the line optical depth). For the other sources, methanol lines are mostly optically thick (FIR6c-a: $\tau_L=$[1.1 $-$ 5.2], MMS9-a: $\tau_L=$[1.2 $-$ 5.7], MMS5: $\tau_L=$[0.9 $-$ 4.2]. The derived gas temperature and density are $\geq$ 85 K and $\geq 3\times 10^6$ cm$^{-3}$ for all sources, with the highest gas density for CSO33-b-a and the lowest gas density for FIR6c-a. The highest gas temperature is derived towards MMS9-a ($\geq$ 130 K). The observed lines are predicted to be emitted by sources between 0.07 and 0.6$''$ ($\sim$ 28 - 236 au) in diameter. Figure \ref{fig:LVG_res} shows as an example the result of the LVG fit for MMS5. The best-fit solutions and ranges obtained for each source are reported in Table \ref{tab:results_LVG}.

\subsection{LTE versus non-LTE analysis}
We provide the results we obtained with the rotational diagram method (LTE) using the same lines as in the LVG analysis in Table~\ref{tab:results_LVG}, in Figure \ref{fig:RDs}. Depending on the sources, the LTE and non-LTE analyses can give similar or different results. In the cases of FIR6c-a and MMS9, the column densities can differ by up to two orders of magnitude. However, this is because, for these sources, we did not know a priori the size of the emitting region and we thus used the sizes from \citet{bouvier2021} which happened to be larger (up to $\sim$40\%) than those we derived with the LVG analysis. Additionally,  we see that the lines in these sources are optically thick. In general, the optical depth and the source size can be corrected using the population diagram method \citep[][]{goldsmith1999}. However, a population diagram cannot correct for non-LTE effects if they are present. 

For each transition line, the excitation temperature corresponding to the best fit of the LVG analysis is indicated in Table B.1. Comparing with the kinetic temperatures derived in the LVG analysis, we can see that some lines are sub-thermally populated and that there are maser lines at 218440 and 261805 MHz. We note that for CSO33-a, where the lines are optically thin and under LTE conditions, we find consistent results between the LTE and LVG analyses. For the source FIR6c-a, for which the excitation temperatures are very different from the derived kinetic temperature, we checked that non-LTE effects remain present even after correcting the rotational diagram for the size and the optical depth (there is still a scatter of points). In other words and as expected, the population diagram method can give a good approximation of the results if the lines are close to being thermally populated, which is only known when a non-LTE analysis is carried-out.

\begin{table*}[ht]
   \caption{Source properties, LTE results, best fit results and 1$\sigma$ confidence level (range) from the Non-LTE LVG analysis, and derived methanol abundances with respect to H$_2$.}
    \label{tab:results_LVG}
    \begin{tabular}{lccccc}
    \hline \hline
        &CSO33-b-a & FIR6c-a & MMS9-a & MMS5 & SIMBA-a  \\
        \hline
        \multicolumn{6}{c}{Source properties\tablenotemark{a}} \\
        \hline
        source size [$''\times''$] & $0.6 \times 0.6$ & 0.31 $\times$ 0.13 & 0.44 $\times 0.14$\tablenotemark{b} &0.15 $\times 0.13$ &0.13 $\times$ 0.11  \\
        (envelope + disk) mass [$\times 10^{-2}$\msol] & $\geq 0.2$ & $1.5 - 4$& 2 $-$ 7 & $1 - 2$ & $1 - 3$ \\
        $T_d$ [K] & 10 $-$ 200 & 89 $-$ 134 & 80 $-$ 200 & 149 $-$ 159 & 160 $-$ 200\\
        H$_2$ [$\times 10^{24}$cm$^{-2}$] & $\geq 0.08$ & 7 $-$ 15 & 5 $-$ 19 & 10 $-$ 15 & 18 $-$ 36 \\
        \hline
        \multicolumn{6}{c}{LTE results}\\
        \hline
        size used [$''$]\tablenotemark{c} &0.6 &0.2 &0.25 &0.14 &0.12\\
        $T_{\text{rot}}$ [K] & 124 $\pm$ 262 &169 $\pm$ 54 & 142 $\pm$ 22 &117 $\pm$ 14 & 151 $\pm$ 598 \\
        $N_{\text{tot}}$ [$\times 10^{15}$cm$^{-2}$]& 13 $\pm$ 11 &21 $\pm$ 6 &48 $\pm$ 7  & 150 $\pm$ 20 &20 $\pm$ 30  \\
        \hline
        \multicolumn{6}{c}{LVG results}\\
        \hline
        $n_{\text{H2}}$ [$\times 10^7 $cm$^{-3}$] best fit  & 300 & 0.4 & 0.7& 5 &1.5 \\
        $n_{\text{H2}}$ [$\times 10^7 $cm$^{-3}$] range & $\geq$20 & 0.3 $-$ 0.5 & 0.6 $-$ 1 &  2 $-$ 20 & $\geq 0.7$ \\
        $T_{\text{kin}}$ [K] best fit &105 & 180 & 170 & 105 & 190\\
        $T_{\text{kin}}$ [K] range & 95$-$ 120  & $\geq 85$ &$\geq$130 & 90 $-$ 125 & $\geq$ 100\\
        $N_{\text{CH3OH}}$ [$\times 10^{16}$cm$^{-2}$] best fit & 1.4 & 120 & 400 & 200 & 0.8\\
        $N_{\text{CH3OH}}$ [$\times 10^{16}$cm$^{-2}$] range & 0.7 $-$ 16 & 80 $-$ 200 & 200 $-$ 600 & 140 $-$ 800 & 0.1 $-$ 3\\
        size [$''$] best fit &0.39 & 0.1 &0.12 & 0.2 & 0.17 \\
        size [$''$] range & 0.1 $-$ 0.6 & 0.07 $-$0.13 & 0.1 $-$ 0.13 & 0.13 $-$ 0.24 & 0.08 $-$ 0.38 \\
        X(CH$_3$OH) $\times 10^{-8}$\tablenotemark{d} & $\leq$200 & 5.3 $-$ 29\tablenotemark{*} & 10 $-$ 120\tablenotemark{*} & 9.3 $-$ 80 & 0.003 $-$ 0.2 \\
         \hline
    \end{tabular}
       \tablenotetext{a}{Derived from a continuum analysis in \citet{bouvier2021}.}
    \tablenotetext{b}{Derived from a continuum analysis in \citet{tobin2020}.}
    \tablenotetext{c}{The size is calculated using the formula $\sqrt{a \times b}$, where $a$ and $b$ are the major and minor axes of the source size derived in \citet{bouvier2021}. }
    \tablenotetext{d}{The H$_2$ column densities can be underestimated when the source size is larger than the region of emission of methanol. The abundances derived in this work should then be taken as upper limits in these cases.}
    \tablenotetext{*}{The methanol abundances are likely upper limits, as the source sizes used to derive the H$_2$ column densities are larger than the methanol emission sizes derived in the LVG analysis. }
\end{table*}

\subsection{Derivation of Methanol Abundances}\label{sec:abundance}

In the previous ORANGES study, we focused on the continuum analysis of the sources \citep{bouvier2021}. We used the spectral energy distribution (SED) method to constrain several dust parameters such as the optical depth, the temperature, the H$_2$ column density and the (envelope+disk) mass. These parameters were estimated for a source size derived from a fit in the visibility plane and are reported in Table~\ref{tab:results_LVG} with the associated source size.

We therefore used these H$_2$ column densities to derive the methanol abundance with respect to H$_2$, X(CH$_3$OH), towards each of the 5 sources. The results are reported in Table \ref{tab:results_LVG}. However, since the source size derived from \citet{bouvier2021} can be larger (up to $\sim$ 40\%) than the size of the methanol emission derived from the LVG analysis, the H$_2$ column densities can be thus underestimated in some cases, and the derived abundances would then need to be taken as upper limits. The abundances range between 3$\times10^{-11}$ and 2$\times 10^{-6}$. For CSO33-b-a, only a lower limit could be derived for the H$_2$ column density, so the methanol abundance derived here is an upper limit. SIMBA-a seems to have a lower methanol abundance than the other sources but since the LVG analysis has been performed with only a few data points for most of the sources, the accuracy of the fit is not very elevated.

\section{Discussion} \label{sec:discussion}

\subsection{New Hot Corinos Discovered in the OMC-2/3 Filament}

So far, only three hot corinos have been identified in the OMC-2/3 filament, the intermediate mass protostars HOPS-87 and HOPS-370 \citep{hsu2020, tobin2019}, and HOPS-108 \citep{tobin2019, chahine2022}. One of the questions we aim to answer is: how many hot corinos are present in the OMC-2/3 filament? 
\begin{figure*}
    \centering
    \includegraphics[width=\linewidth]{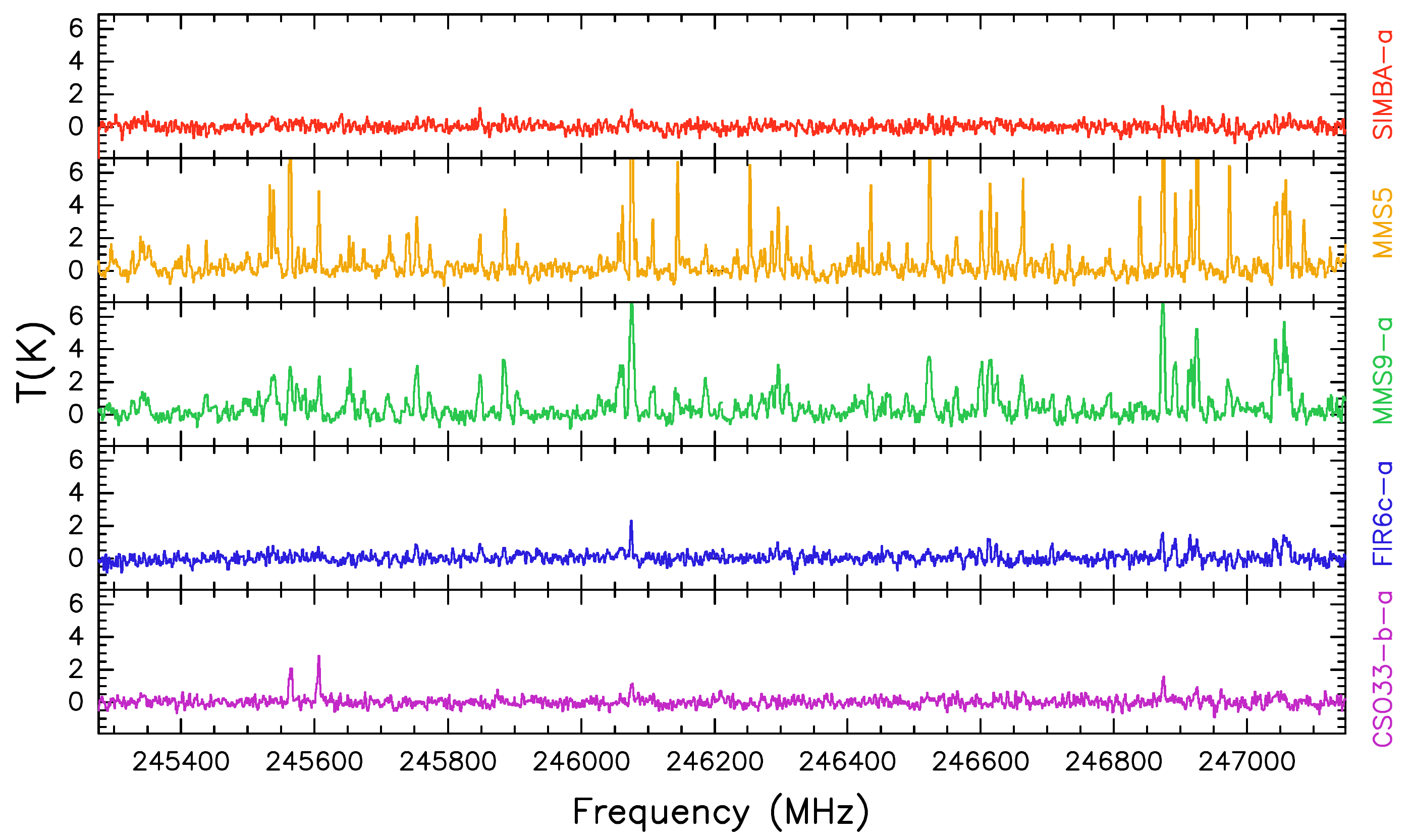}
    \caption{Spectra of each source from the large spectral window, setup 1. The spectra is extracted from a pixel at the peak of the emission.}
    \label{fig:spectra-hc}
\end{figure*}

Our results show that methanol is detected towards 5 protostars from our source sample and that the emission comes from a hot ($\geq$ 85 K), dense ($ \geq 3\times 10^6$ cm$^{-3}$) and compact (0.1 $-$ 0.6$''$ or $\sim$ 39 $-$ 236 au) region. According to the hot corino definition, i.e. a compact ($\leq 100$ au), hot ($\geq 100$ K), and dense ($\geq$ 10$^7$ cm$^{-3}$) region enriched in iCOMs \citep{ceccarelli2004, ceccarelli2007}, CSO33-b-a, FIR6c-a, MMS9-a, MMS5 and SIMBA-a are, therefore, bona fide hot corinos.\footnote{In this work, we targeted only CH$_3$OH which is the most abundant iCOMs found in hot corinos. Other iCOMs could be also present but their identification will be the subject of a future work.} The methanol abundances derived towards the OMC-2/3 hot corinos are comparable to what is derived in other hot corinos in Orion (HOPS-87, HOPS-168, HOPS-288, G192.12-11.10, HH212; \citealt{lee2019, hsu2020}) and in other star-forming regions (e.g. B335, IRAS16293-2422; \citealt{imai2016, jorgensen2016, jorgensen2018}), except for SIMBA-a for which the methanol abundance is about 2 orders of magnitude lower. However, for FIR6c-a and MMS9-a, the abundances could be overestimated (see Sec.~\ref{sec:abundance}), and most of the LVG analyses were performed with only three lines. Our results should thus be taken with caution.

The five hot corinos show very different spectra as shown in Fig. \ref{fig:spectra-hc}. MMS5 and MMS9-a present line-rich spectra with strong iCOM emission whilst CSO33-b-a, FIR6c-a and SIMBA-a present line-poor spectra, likely because the iCOM emission is faint. We will address the analysis of the other iCOMs detected towards the sources in a forthcoming paper.

\subsection{Is the Dust Hiding Other Hot Corinos?}\label{sec:dust}

A recent study by \citet{de_simone2020} showed that hot corinos detected at centimetre wavelengths could be obscured by optically thick dust at millimetre wavelengths. Could it be the case for some of our sources?  

\begin{figure*}[ht]
    \centering
    \includegraphics[width=0.8\linewidth]{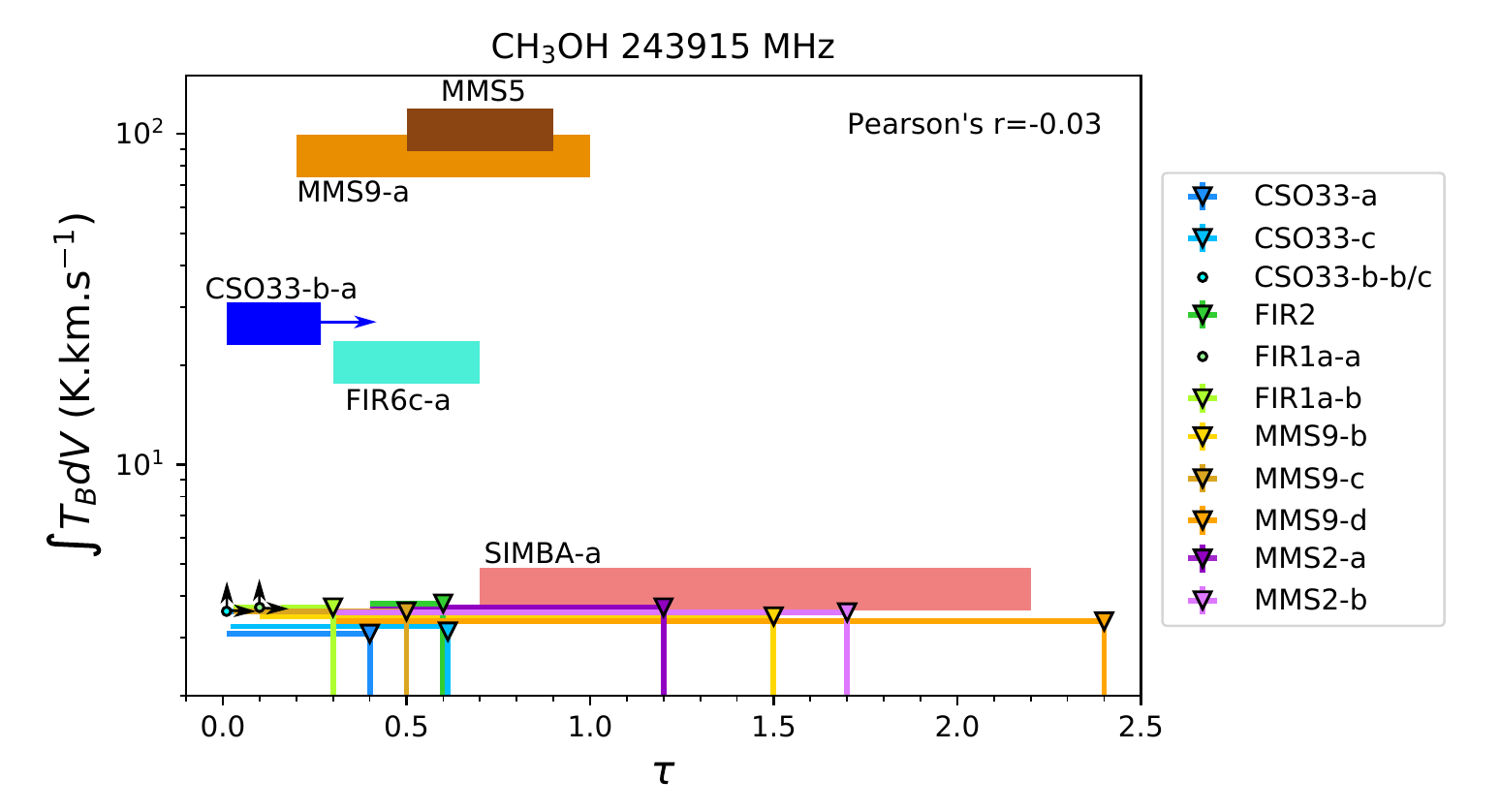}
    \caption{Line intensity of the CH$_3$OH line at 243915 MHz as a function of the dust optical depth, $\tau$. For clarity, we slightly shifted vertically the upper limits for the line intensity of several sources. The initial upper limit for the CH$_3$OH line is 3.6 K.km.s$^{-1}$ for the components of the systems CSO33 and MMS9, 3.7 K.km.s$^{-1}$ for the FIR1a and MMS2 components and for CSO3-b, and 3.8 K.km.s$^{-1}$ for FIR2. Upper limits are represented by coloured filled triangles or arrows.}
    \label{fig:int-tau}
\end{figure*}

Figure \ref{fig:int-tau} shows the line intensity of the CH$_3$OH transition line at 243915 MHz as a function of the dust opacity. The latter has been derived for each source of the sample in \citet{bouvier2021}. For sources where no methanol is detected, we calculated the 3$\sigma$ upper limit for the line intensity. If the optical depth was a dominant factor, we would expect to see an anti-correlation between the methanol intensity and $\tau$, with the sources presenting methanol lines having the lowest range of dust optical depths. We do not see any anti-correlation which suggests that the dust opacity is not the main parameter affecting the detection of methanol, and hence the detection of hot corinos, in the OMC-2/3 filament. However, we note that the dust optical depth ranges derived in \citet{bouvier2021} do not always correspond to the sizes derived from the LVG analysis performed in this work. In some cases (FIR6c-a and MMS9-a), we derived methanol emission sizes that are smaller than the size of the continuum emission. This would indicate that we are underestimating the dust optical depth at the scale probed by the methanol emission. Therefore, our conclusion needs to be taken with caution. Additionally, we can see that for four of our sample sources (MMS2-a, MMS2-b, MMS9-b, and MMS9-d), the upper limits for the derived dust optical depths are larger than 1. In these sources, we, thus, cannot exclude the possibility that the dust absorbs methanol emission at 1.3mm.

\subsection{Are ORANGES Different From PEACHES?}

Several studies targeting methanol and other iCOMs towards low-mass protostars have been conducted. \citet{yang2021} surveyed 50 sources in the Perseus Molecular cloud in the context of PEACHES. They detected CH$_3$OH towards 56\% of their source sample and other O-bearing iCOMs towards 32\% of the source sample. \citet{belloche2020} surveyed 16 class 0 protostars located in various low-mass star-forming regions as part of the Continuum And Lines in Young ProtoStellar Objects (CALYPSO) IRAM Large Program survey, with the Plateau de Bure Interferometer (PdBI, the predecessor of the current NOEMA interferometer). They detected methanol emission towards 50\% of their source sample, but no more than 30\% of them with at least three iCOMs detected. \citet{vangelder2020} (ALMA) surveyed 7 Class 0 sources in the Perseus and Serpens molecular clouds and detected methanol towards three of them ($\sim 43$\%). Finally, \citet{bergner2017} IRAM-30m targeted iCOMs towards 16 Class 0/I protostars and detected the iCOMs CH$_3$CHO, CH$_3$OCH$_3$, and CH$_3$OCHO towards 37\%, 13\%, and 13\% of the sources, respectively. However, contrarily to the other surveys cited above, the temperatures derived by \citet{bergner2017}  being too low ($\leq 30$ K) for the iCOMs to originate from a hot corino region, the emission of iCOMs could trace a more external component. These surveys show that selecting a mix of usual targets, methanol is largely detected in solar-mass protostars located in low-mass star-forming regions. Here, we compare our results with those of PEACHES only, as this is the only unbiased survey targeting iCOMs towards all the protostars of a single low-mass star-forming region. Additionally, the PEACHES and ORANGES were designed to compare directly the low-mass protostellar chemical content of two different environments, the Perseus Molecular Cloud and the OMC-2/3 filament. In both regions, the selected targets are mostly Class 0, I, or 0/I protostars with a low fraction of other (Class II or unknown) sources (7\% and 11\% of the sources in PEACHES and ORANGES, respectively). The relative fraction of Class 0 and I sources in each region cannot be determined accurately as the current classification of the protostars is either based on Herschel observations, for which the angular resolution is not sufficient to disentangle close multiple systems, or not certain. However, as hot corinos are detected both towards Class 0 and I sources, this parameter is not particularly relevant when comparing the two regions. Finally, the distances of the two clouds have been taken into account to achieve the same sensitivity ($\sim$22 mJy/beam for PEACHES and $\sim$24 mJy/beam for ORANGES) and spatial resolution for the two projects. 

The results from PEACHES showed that (56 $\pm$ 14)\% of their source sample present warm methanol emission \citep{yang2021}, which means that bona fide hot corinos are common in the Perseus Molecular Cloud. On the other hand, we targeted 19 solar-mass protostars located in the OMC-2/3 filament and detected only five bona fide hot corinos. Even though three other hot corinos are located in the OMC-2/3 filament (HOPS-87, HOPS-370, HOPS-108; \citealt{tobin2019, hsu2020, chahine2022}) we do not take them into account. Indeed, unlike the above-cited studies, we performed a blind search for hot corinos using an observational setup completely analogous to that done in Perseus by \citet{yang2021}. We, thus, do not want to bias our results by adding only positive hot corino detections from other studies. Finally, using only our results, we have a hot corino detection rate of (26 $\pm$ 23)\% in the OMC-2/3 filament. Therefore, hot corinos seem to be scarcer in the OMC-2/3 filament, compared to the Perseus Molecular Cloud.

The two star-forming regions seem to have different chemical protostellar content but the high uncertainty for the ORANGES survey prevents us from firmly concluding. We need to increase the statistics and to do so, a possibility would be to include the Class 0 and I population of the OMC-4 cloud, located south of OMC-1. Additionally, as mentioned in Sec.~\ref{sec:dust}, we cannot exclude that dust could hide hot corinos towards some of our source sample and, therefore, that the detection rate of 26\% is underestimated. If there is truly a difference between ORANGES and PEACHES, then the environment most likely plays a role in shaping the chemical content of protostellar cores. Bounded by 3 HII regions, the OMC-2/3 filament is highly illuminated by ultraviolet photons, and if hot corinos are less abundant in this kind of region, it would be in line with recent modelling and observational studies \citep[][]{aikawa2020,lattanzi2020,kalvans2021}: a cloud exposed to interstellar irradiation is very likely to be less rich in O-bearing species and in iCOMs than a more shielded one. 

Although we cannot totally dismiss the possibilities that (1) some of our protostars may have small hot corino regions, preventing us from detecting iCOMs at our current resolution, (2) high dust optical depths could still play a role in the non-detection of hot corinos in some of our sample sources, this study provides tentative evidence of a differentiation of the chemical nature of solar-mass protostars that are located in two different environments or, in other words, that ORANGES may be different from PEACHES.

%% rewrite this sentence?

\section{Conclusion}

The ORion ALMA New GEneration Survey aims to study the small-scale ($\leq 100$ au) chemical content of solar-mass protostars located in the  highly illuminated OMC-2/3 filament. We detected methanol emission centred towards 5 out of the 19 targeted sources. After performing a non-LTE LVG analysis, we showed that the methanol-emitting regions are hot (T$\geq$ 85 K), dense ($n_{\text{H2}}\geq 3.10^6$ cm$^{-2}$) and compact ($\sim$ 0.1 $-$ 0.6$''$ or $\sim$ 39 $-$ 236 au in diameter), and correspond to hot corino regions. We thus detected five new bona fide hot corinos in the OMC-2/3 filament, which corresponds to (26 $\pm$ 23)\% of the sample sources.

%We also showed that in our case, performing a non-LTE LVG analysis give more robust result crucial to derive robust parameters as there are optically thick lines, masers, and sub-thermally populated lines.

On the other hand, a similar study performed in the less illuminated low-mass star-forming region of Perseus found a high detection rate, (56 $\pm$ 14) \%, of hot corinos \citep{yang2021}. Hot corinos seem thus scarcer in a highly illuminated environment such as the OMC-2/3 filament. This result indicates that the environment may very likely playing a role in solar-mass protostars chemical content and that ORANGES are different from PEACHES.

Are hot corinos always abundant in low-mass star-forming regions analogue to Perseus and more scarce in analogues to the OMC-2/3 filament? We would need to perform more studies analogous to PEACHES and ORANGES in other star-forming regions to confirm this result. Finally, although hot corinos are present in a region similar to the one in which our Sun is born, they are not prevailing. The question of whether our Sun experienced a hot corino phase in its youth needs further investigations before being answered. \\

%%%%%%%%%%% Aknowledgments %%%%%%%%%%%%%
\textit{Acknowledgments.}
We deeply thank the anonymous referee for their helpful comments that contributed to significantly improving the paper. While the paper was under review, three additional hot corinos were detected in the OMC-2/3 filament (HOPS-84-A, HOPS-84-B, and MMS1) by \citet{hsu2022}. Moreover, they targeted 56 Class 0/I protostars throughout the Orion Molecular cloud and detected warm methanol towards ~20\% of their sample sources, which is comparable to what we found in this work.
This project has received funding from the European Research Council (ERC) under the European Union's Horizon 2020 research and innovation programme, for the Project \textit{The Dawn of Organic Chemistry} (DOC), grant agreement No 741002. This paper makes use of the following ALMA data: ADS/JAO.ALMA\#2016.1.00376.S. ALMA is a partnership of ESO (representing its member states), NSF (USA) and NINS (Japan), together with NRC (Canada) and NSC and ASIAA (Taiwan), in cooperation with the Republic of Chile. The Joint ALMA Observatory is operated by ESO, AUI/NRAO and NAOJ.


\bibliography{refs}{}

\begin{thebibliography}{}
\expandafter\ifx\csname natexlab\endcsname\relax\def\natexlab#1{#1}\fi
\providecommand{\url}[1]{\href{#1}{#1}}
\providecommand{\dodoi}[1]{doi:~\href{http://doi.org/#1}{\nolinkurl{#1}}}
\providecommand{\doeprint}[1]{\href{http://ascl.net/#1}{\nolinkurl{http://ascl.net/#1}}}
\providecommand{\doarXiv}[1]{\href{https://arxiv.org/abs/#1}{\nolinkurl{https://arxiv.org/abs/#1}}}

\bibitem[{Adams(2010)}]{adams2010}
Adams, F.~C. 2010, ARA\&A, 48, 47

\bibitem[{Aikawa {et~al.}(2020)Aikawa, Furuya, Yamamoto, \& Sakai}]{aikawa2020}
Aikawa, Y., Furuya, K., Yamamoto, S., \& Sakai, N. 2020, ApJ, 897, 110

\bibitem[{Belloche {et~al.}(2020)Belloche, Maury, Maret,
  {et~al.}}]{belloche2020}
Belloche, A., Maury, A.~J., Maret, S., {et~al.} 2020, A\&A, 635, A198

\bibitem[{Bergner {et~al.}(2017)Bergner, \"Oberg, Garrod, \&
  Graninger}]{bergner2017}
Bergner, J.~B., \"Oberg, K.~I., Garrod, R.~T., \& Graninger, D.~M. 2017, ApJ,
  841, 120

\bibitem[{Bianchi {et~al.}(2020)Bianchi, Chandler, Ceccarelli,
  {et~al.}}]{bianchi2020}
Bianchi, E., Chandler, C.~J., Ceccarelli, C., {et~al.} 2020, MNRAS, 498, L87

\bibitem[{Bianchi {et~al.}(2019)Bianchi, Codella, Ceccarelli,
  {et~al.}}]{bianchi2019}
Bianchi, E., Codella, C., Ceccarelli, C., {et~al.} 2019, MNRAS, 483, 1850

\bibitem[{Bouvier {et~al.}(2020)Bouvier, L\'opez-Sepulcre, Ceccarelli, Kahane,
  Imai, Sakai, Yamamoto, \& Dagdigian}]{bouvier2020}
Bouvier, M., L\'opez-Sepulcre, A., Ceccarelli, C., {et~al.} 2020, A\&A, 636,
  A19

\bibitem[{Bouvier {et~al.}(2021)Bouvier, L\'opez-Sepulcre, Ceccarelli, Sakai,
  Yamamoto, \& Yang}]{bouvier2021}
---. 2021, A\&A, 653, A117

\bibitem[{Ceccarelli(2004)}]{ceccarelli2004}
Ceccarelli, C. 2004, Astronomical Society of the Pacific Conference Series,
  323, 195

\bibitem[{Ceccarelli {et~al.}(2007)Ceccarelli, Caselli, Herbst, Tielens, \&
  {Caux}}]{ceccarelli2007}
Ceccarelli, C., Caselli, P., Herbst, E., Tielens, A. G. G.~M., \& {Caux}, E.
  2007, Protostars and Planets V, B. Reipurth, D. Jewitt and K. Keil (eds.),
  University of Arizona Press, 47

\bibitem[{Ceccarelli {et~al.}(2003)Ceccarelli, Maret, Tielens, Castets, \&
  Caux}]{ceccarelli2003}
Ceccarelli, C., Maret, S., Tielens, A. G. G.~M., Castets, A., \& Caux, E. 2003,
  A\&A, 410, 587

\bibitem[{Ceccarelli {et~al.}(2017)Ceccarelli, Caselli, Fontani, Neri,
  L\'opez-Sepulcre, Codella, Feng, Jimenez-Serra, Lefloch, Pineda, Vastel,
  Alves, Bachiller, Balucani, Bianchi, Bizzochi, Bottinelli, Caux, \&
  Chacon-Tanarro}]{ceccarelli2017}
Ceccarelli, C., Caselli, P., Fontani, F., {et~al.} 2017, ApJ, 850, 176

\bibitem[{Chahine {et~al.}(2022)Chahine, L\'{o}pez-Sepulcre, Neri,
  {et~al.}}]{chahine2022}
Chahine, L., L\'{o}pez-Sepulcre, A., Neri, R., {et~al.} 2022, A\&A, 657, A78

\bibitem[{Chini {et~al.}(1997)Chini, Reipurth, Ward-Thompson, Bally, Nyman,
  Sievers, \& Billawala}]{chini1997}
Chini, R., Reipurth, B., Ward-Thompson, D., {et~al.} 1997, ApJ, 474, L135

\bibitem[{Codella {et~al.}(2016)Codella, Ceccarelli, Cabrit, Gueth, Podio,
  Bachiller, Fontani, Gusdorf, Lefloch, Leurini, \& Tafalla}]{codella2016}
Codella, C., Ceccarelli, C., Cabrit, S., {et~al.} 2016, A\&A, 586, L3

\bibitem[{{de Jong} {et~al.}(1980){de Jong}, Boland, \& Dalgarno}]{dejong1980}
{de Jong}, T., Boland, W., \& Dalgarno, A. 1980, A\&A, 91, 68

\bibitem[{De~Simone {et~al.}(2020)De~Simone, Codella, Ceccarelli,
  {et~al.}}]{de_simone2020}
De~Simone, M., Codella, C., Ceccarelli, C., {et~al.} 2020, ApJL, 896, L3

\bibitem[{Drozdovskaya {et~al.}(2019)Drozdovskaya, van Dishoeck, Rubin,
  J\o{rgensen}, \& Altwegg}]{drozdovskaya2019}
Drozdovskaya, M.~N., van Dishoeck, E.~F., Rubin, M., J\o{rgensen}, J.~K., \&
  Altwegg, K. 2019, MNRAS, 490, 50

\bibitem[{Dubernet {et~al.}(2013)Dubernet, Alexander, Ba, Balakrishnan,
  Balança, Ceccarelli, Cernicharo, Daniel, Dayou, Doronin, Dumouchel, Faure,
  Feautrier, Flower, Grosjean, Halvick, Klos, Lique, McBane, Marinakis, Moreau,
  Moszynski, Neufeld, Roueff, Schilke, Spielfiedel, Stancil, Stoecklin,
  Tennyson, Yang, Vasserot, \& Wiesenfeld}]{dubernet2013}
Dubernet, M.-L., Alexander, M.~H., Ba, Y.~A., {et~al.} 2013, A\&A, 553, A50

\bibitem[{Feddersen {et~al.}(2020)Feddersen, Arce, Kong,
  {et~al.}}]{feddersen2020}
Feddersen, J.~R., Arce, H.~G., Kong, S., {et~al.} 2020, ApJ, 896, 11

\bibitem[{Fischer {et~al.}(2013)Fischer, Megeath, Stutz, Tobin, Ali, Stanke,
  {et~al.}}]{fischer2013}
Fischer, W.~J., Megeath, S.~T., Stutz, A.~M., {et~al.} 2013, AN, 334, 53

\bibitem[{Fisher {et~al.}(2007)Fisher, Paciga, Xu, Zhao, Moruzzi, \&
  Lees}]{fisher2007}
Fisher, J., Paciga, G., Xu, L.-H., {et~al.} 2007, J. Mol. Spectrosc. 245, 7

\bibitem[{Flower {et~al.}(2010)Flower, Pineau~des For\^{e}ts, \&
  Rabli}]{flower2010}
Flower, D.~R., Pineau~des For\^{e}ts, G., \& Rabli, D. 2010, MNRAS, 409, 29

\bibitem[{Furlan {et~al.}(2016)Furlan, Fischer, Ali, Stutz, Stanke, Tobin,
  Megeath, {et~al.}}]{furlan2016}
Furlan, E., Fischer, W.~J., Ali, B., {et~al.} 2016, ApJS, 224, 5

\bibitem[{Goldsmith \& Langer(1999)}]{goldsmith1999}
Goldsmith, P.~F., \& Langer, W.~D. 1999, ApJ, 517, 209

\bibitem[{G\'omez-Ruiz {et~al.}(2019)G\'omez-Ruiz, Gusdorf, Leurini,
  {et~al.}}]{gomezruiz2019}
G\'omez-Ruiz, A.~I., Gusdorf, A., Leurini, S., {et~al.} 2019, A\&A, 629, 77

\bibitem[{Gro{\ss}schedl {et~al.}(2018)Gro{\ss}schedl, Alves, Meingast, Ackerl,
  Ascenso, Bouy, Burkert, Forbrich, Fuernkranz, Goodman, Hacar, Herbst-Kiss,
  Lada, Larreina, Leschinski, Lombardi, Moitinho, Mortimer, \&
  Zari}]{grosschedl2018}
Gro{\ss}schedl, J.~E., Alves, J., Meingast, S., {et~al.} 2018, A\&A, 619, A106

\bibitem[{Herbst \& Van~Dishoeck(2009)}]{herbst2009}
Herbst, E., \& Van~Dishoeck, E.~F. 2009, ARA\&A, 47, 427

\bibitem[{Hsu {et~al.}(2020)Hsu, Liu, Liu, {et~al.}}]{hsu2020}
Hsu, S.-Y., Liu, S.-Y., Liu, T., {et~al.} 2020, ApJ, 898, 107

\bibitem[{Hsu {et~al.}(2022)Hsu, Liu, Liu, {et~al.}}]{hsu2022}
---. 2022, ApJ, in press

\bibitem[{Imai {et~al.}(2016)Imai, Sakai, Oya, L\'opez-Sepulcre, Watanabe,
  Ceccarelli, Lefloch, Caux, Vastel, Kahane, Sakai, Hirota, Aikawa, \&
  Yamamoto}]{imai2016}
Imai, M., Sakai, N., Oya, Y., {et~al.} 2016, ApJ, 830, L37

\bibitem[{Jacobsen {et~al.}(2019)Jacobsen, J\o{rgensen}, {Di Francesco},
  {et~al.}}]{jacobsen2019}
Jacobsen, S.~K., J\o{rgensen}, J.~K., {Di Francesco}, J., {et~al.} 2019, A\&A,
  629, A29

\bibitem[{J\o{rgensen} {et~al.}(2018)J\o{rgensen}, M\"uller, Calcutt,
  {et~al.}}]{jorgensen2018}
J\o{rgensen}, J.~K., M\"uller, H. S.~P., Calcutt, H., {et~al.} 2018, A\&A, 620,
  A170

\bibitem[{J\o{rgensen} {et~al.}(2016)J\o{rgensen}, van~der Wiel, Coutens,
  Lykke, Müller, van Dishoeck, Calcutt, Bjerkeli, Bourke, Drozdovskaya, Favre,
  Fayolle, Garrod, Jacobsen, \"Oberg, Persson, \& Wampfler}]{jorgensen2016}
J\o{rgensen}, J.~K., van~der Wiel, M. H.~D., Coutens, A., {et~al.} 2016, A\&A,
  595, A117

\bibitem[{Kalvans(2021)}]{kalvans2021}
Kalvans, J. 2021, ApJ, 910, 54

\bibitem[{Lattanzi {et~al.}(2020)Lattanzi, Bizzocchi, Vasyunin,
  {et~al.}}]{lattanzi2020}
Lattanzi, V., Bizzocchi, L., Vasyunin, A.~I., {et~al.} 2020, A\&A, 633, A118

\bibitem[{Lee {et~al.}(2019)Lee, Codella, Li, \& Liu}]{lee2019}
Lee, C.-F., Codella, C., Li, Z.-Y., \& Liu, S.-Y. 2019, ApJ, 876, 63

\bibitem[{Lis {et~al.}(1998)Lis, Serabyn, Keene, Dowell, Benford, Phillips,
  Hunter, \& Wang}]{lis1998}
Lis, D.~C., Serabyn, E., Keene, J., {et~al.} 1998, ApJ, 509, 299

\bibitem[{Matsushita {et~al.}(2019)Matsushita, Takahashi, Machida, \&
  Tomisaka}]{matsushita2019}
Matsushita, Y., Takahashi, S., Machida, M.~N., \& Tomisaka, K. 2019, ApJ, 871,
  221

\bibitem[{McMullin {et~al.}(2007)McMullin, Waters, Schiebel, Young, \&
  Golap}]{mccmullin2007}
McMullin, J.~P., Waters, B., Schiebel, D., Young, W., \& Golap, K. 2007, ASPC,
  376, 127

\bibitem[{M\"{u}ller {et~al.}(2005)M\"{u}ller, Schl\"oder, Stutzki, \&
  Winnewisser}]{muller2005}
M\"{u}ller, H. S.~P., Schl\"oder, F., Stutzki, J., \& Winnewisser, G. 2005,
  JMoSt, 742, 215

\bibitem[{Nielbock {et~al.}(2003)Nielbock, Chini, \& M\"{u}ller}]{nielbock2003}
Nielbock, M., Chini, R., \& M\"{u}ller, S. A.~H. 2003, A\&A, 408, 245

\bibitem[{Oya {et~al.}(2017)Oya, Sakai, Watanabe, Higuchi, Hirota,
  L\'opez-Sepulcre, Sakai, Aikawa, Ceccarelli, Lefloch, Caux, Vastel, Kahane,
  \& Yamamoto}]{oya2017}
Oya, Y., Sakai, N., Watanabe, Y., {et~al.} 2017, ApJ, 837, 174

\bibitem[{Pfalzner {et~al.}(2015)Pfalzner, Davies, Gounelle, Johansen,
  M\"unker, Lacerda, Portegies~Zwart, Testi, Trieloff, \& Veras}]{pfalzner2015}
Pfalzner, S., Davies, M.~B., Gounelle, M., {et~al.} 2015, PhyS, 90, 068001

\bibitem[{Rivilla {et~al.}(2020)Rivilla, Drodovskaya, Altwegg,
  {et~al.}}]{rivilla2020}
Rivilla, V.~M., Drodovskaya, M.~N., Altwegg, K., {et~al.} 2020, MNRAS, 492,
  1180

\bibitem[{Sakai {et~al.}(2008)Sakai, Sakai, Hirota, \& Yamamoto}]{sakai2008b}
Sakai, N., Sakai, T., Hirota, T., \& Yamamoto, S. 2008, ApJ, 672, 371

\bibitem[{Sakai \& Yamamoto(2013)}]{sakai2013}
Sakai, N., \& Yamamoto, S. 2013, ChRv, 113, 8981

\bibitem[{Shimajiri {et~al.}(2009)Shimajiri, Takahashi, Takakuwa, Saito, \&
  Kawabe}]{shimajiri2009}
Shimajiri, Y., Takahashi, S., Takakuwa, S., Saito, M., \& Kawabe, R. 2009,
  PASJ, 61, 1055

\bibitem[{Takahashi {et~al.}(2008)Takahashi, Saito, Ohashi, Kasakabe, Takakuwa,
  Shimajiri, Tamura, \& Kawabe}]{takahashi2008}
Takahashi, S., Saito, M., Ohashi, N., {et~al.} 2008, ApJ, 688, 344

\bibitem[{Tanabe {et~al.}(2019)Tanabe, Nakamura, Tsukagoshi,
  {et~al.}}]{tanabe2019}
Tanabe, Y., Nakamura, F., Tsukagoshi, T., {et~al.} 2019, PASJ, 71, 8

\bibitem[{Tobin {et~al.}(2019)Tobin, Megeath, van't Hoff, D\'iaz-Rodr\'iguez,
  Reynolds, Osorio, {et~al.}}]{tobin2019}
Tobin, J.~J., Megeath, S.~T., van't Hoff, M., {et~al.} 2019, ApJ, 886, 6

\bibitem[{Tobin {et~al.}(2020)Tobin, Sheehan, Megeath, D\'iaz-Rodr\'iguez,
  Offner, Murillo, {et~al.}}]{tobin2020}
Tobin, J.~J., Sheehan, P.~D., Megeath, S.~T., {et~al.} 2020, ApJ, 890, 130

\bibitem[{{van Gelder} {et~al.}(2020){van Gelder}, Tabone, Tychoniec,
  {et~al.}}]{vangelder2020}
{van Gelder}, M.~L., Tabone, B., Tychoniec, L., {et~al.} 2020, A\&A, 639, A87

\bibitem[{Williams {et~al.}(2003)Williams, Plambeck, \& Heyer}]{williams2003}
Williams, J.~P., Plambeck, R.~L., \& Heyer, M.~H. 2003, ApJ, 591, 1025

\bibitem[{Wilson \& Rood(1994)}]{WR94}
Wilson, T.~L., \& Rood, R. 1994, ARA\&A, 32, 191

\bibitem[{Xu {et~al.}(2008)Xu, Fischer, Lees, Shi, Hougen, Pearson, Drouin,
  Blake, \& Braakman}]{xu2008}
Xu, L.-H., Fischer, J., Lees, R.~M., {et~al.} 2008, J. Mol. Spectrosc. 251, 305

\bibitem[{Yang {et~al.}(2021)Yang, Sakai, Zhang, {et~al.}}]{yang2021}
Yang, Y.-L., Sakai, N., Zhang, Y., {et~al.} 2021, ApJ, 910, 20

\end{thebibliography}
\bibliographystyle{aasjournal}

\appendix

\section{Observational details}
We present here the details of the observations. Table \ref{tab:sources-sample} lists the targeted sources and their coordinates and Table \ref{tab:methanol-spw} shows the list of methanol transitions detected and used in this work, and their spectral parameters. Channel spacing and primary beam size for the spectral windows containing the methanol lines are also indicated. 

\begin{table}[ht]
\renewcommand\thetable{A.1}
 \caption{Sample sources, coordinates of the dust peak continuum (D), coordinates of the positions selected to extract the spectra (P), source classification, and associated HOPS names.}
    \label{tab:sources-sample}
    \centering
    %\begin{threeparttable}[t]
    \resizebox{19cm}{!}{
    \begin{tabular}{lccccccl}
    \hline\hline
   \multirow{2}{*}{Source}&R.A. (D)  & Dec. (D)  &R.A. (P)&Dec. (P)&\multirow{2}{*}{HOPS name$^{a,b}$}&\multirow{2}{*}{Classification$^{c}$}  &\multirow{2}{*}{Notes}\\ 
   &[J2000]&[J2000]&[J2000]&[J2000]&& \\
    \hline
    CSO33-a &05:35:19.41 &$-$05:15:38.41&...&...&HOPS-56-B  &0 or I&\\
    CSO33-b &05:35:19.48 &$-$05:15:33.08&05:35:19.48&-05:15:33.10&HOPS-56-A-A/B/C  &0&triple system$^{d}$ \\
    CSO33-c & 05:35:19.81&$-$05:15:35.22&...&...&V2358 Ori  &II&\\
    FIR6c-a &05:35:21.36&$-$05:13:17.85  &05:35:21.36&-05:13:17.85&HOPS-409 &0&\\
    FIR2& 05:35:24.30& $-$05:08:30.74 &...&...&HOPS-68  &I&\\
    FIR1a-a &05:35:24.87 &$-$05:07:54.63 &...&...&HOPS-394-B  &0 or I& \\
    FIR1a-b &05:35:24.05 &$-$05:07:52.07 &...&...&HOPS-394-A  &0&\\
     MMS9-a &05:35:25.97 &$-$05:05:43.34 &05:35:25.96&-05:05:43.39&HOPS-78-A  &0&\\
    MMS9-b &05:35:26.15 &$-$05:05:45.80 &...&...&HOPS-78-B &0 or I& \\
    MMS9-c &05:35:26.18&$-$05:05:47.14  &...&...& HOPS-78-C &0 or I&\\
    MMS9-d & 05:35:25.92&$-$05:05:47.70 &...&...& HOPS-78-D &II?&\\
    MMS5 &05:35:22.47 &$-$05:01:14.34 &05:35:22.48&-05:01:14.35& HOPS-88&0&\\
     MMS2-a &05:35:18.34&$-$05:00:32.96  &...&...& HOPS-92-A-A/B &I& binary$^{d}$ \\
    MMS2-b & 05:35:18.27& $-$05:00:33.95&...&...&HOPS-92-B &I&\\
    CSO3-b &05:35:16.17 &$-$05:00:02.50 &...&... &HOPS-94 &I&\\
    SIMBA-a &05:35:29.72 &$-$04:58:48.60&05:35:29.72&-04:58:48.56&HOPS-96  &0&\\
    \hline
    \end{tabular}}
    %\end{threeparttable}
    \begin{minipage}{20cm}
     $^a$\citealt{fischer2013} $^b$\citealt{furlan2016} $^c$\citealt{bouvier2021} $^d$\citealt{tobin2020}
    \end{minipage}
\end{table}

%% add synthesized beams in table below
\begin{table*}[ht]
    \centering
    \renewcommand\thetable{A.2}
       \caption{Methanol transition lines detected in this work, their parameters, and channel spacing and primary beam size of the associated spectral windows. }
    \label{tab:methanol-spw}
    \begin{tabular}{cccccccc}
    \hline \hline
         Molecule & Frequency & Transition & $E_{up}$& $g_{up}$ & $A_{ij}$ & channel spacing & primary beam size \\
         &[MHz] & & [K] & & [$\times 10^{-5}$s$^{-1}$] & [km.s$^{-1}$] &[$''$]\\
         \hline
         CH$_3$OH &218440&$4_{-2,3} - 3_{-1,2} $ E& 45.5 & 36 & 4.69& 0.5 & 28.8 \\
         &232418 &$10_{2,8} - 9_{3,7} $ A &  165.4 & 84 & 1.87 & 1.3 & 27.1 \\ 
         & 232945 &$10_{-3,7} - 11_{-2,9} $ E& 190.4&84&2.13 &1.3 & 27.1\\
         &234683&$4_{2,3} - 5_{1,4} $ A& 60.9 & 36 & 1.87 &0.5 & 26.8\\
         &234698&$5_{4,2} - 6_{3,3} $ E&122.7 & 44 & 0.63 &0.5 & 26.8\\
         &243915 &$5_{1,4} - 4_{1,3} $ A& 49.7 &44 & 5.97 &0.5 & 25.8\\
         &261805 &$2_{1,1} - 1_{0,1} $ E& 28.0 &20 & 5.57 &0.5 & 24.1\\
         \hline
          CH$_3^{18}$OH &231758 &$5_{0,5} - 4_{0,4} $ A& 33.4 &44 &5.33&1.3 &27.1  \\
         % &231.8538& $5_{-2,4} - 4_{-2,3} $ E & 59.2 &44& 4.49 &\\
          %&231.8645&$5_{2,3} - 4_{2,2} $E&55.8&44&4.41&\\
         \hline
    \end{tabular}
 \tablecomments{Frequencies and spectroscopic parameters have been extracted from the CDMS catalogue \citep{muller2005}. For CH$_3$OH (TAG 032504, version 3$^*$) and CH$_3^{18}$OH (TAG 034504, version 1$^*$), the available data are from \citet{xu2008} and \citet{fisher2007}, respectively.}
\end{table*}

\section{Gaussian fit results and CH$_3^{18}$OH spectrum}
\begin{figure}[ht]
    \centering
    \renewcommand\thefigure{B.1}
    \includegraphics[width=0.8\linewidth]{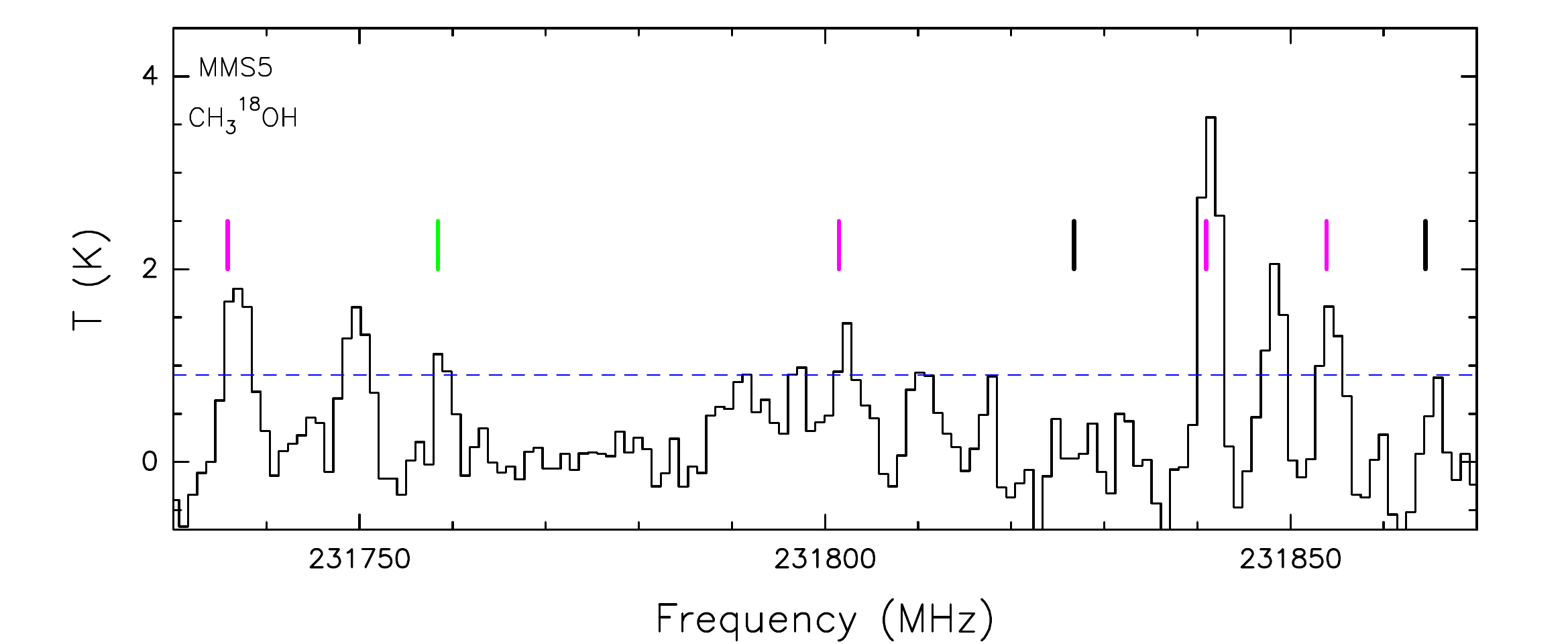}
    \caption{Spectra towards MMS5 where the frequencies of the seven CH$_3^{18}$OH lines expected to be the most intense  ($E_u < 75K$) are indicated. Detected lines are marked in green, contaminated lines are marked in magenta, and undetected lines are marked in black. The 3$\sigma$-level is indicated by the dashed blue line.}
    \label{fig:ch3oh18h}
\end{figure}

The Gaussian fit results of the CH$_3$OH and CH$_3^{18}$OH lines are reported in Table \ref{tab:spectral-params} Contaminated lines are not reported in the table as they are not included into the LVG fit. Figure \ref{fig:ch3oh18h} shows the detected transition of CH$_3^{18}$OH towards MMS5. 

%%move syntesized beams to Table A and put fitted size from continuum
\begin{table*}[ht]
    \centering
    \renewcommand\thetable{B.1}
     \caption{List of frequencies of the detected methanol lines, synthesized beams, and line fitting and LVG results.}
    \label{tab:spectral-params}
    \resizebox{19cm}{!}{
    \begin{tabular}{cccrrrrcccc}
        \hline \hline
    Molecule& Frequency & Synthesized Beam &$\int T_BdV$ G.&$\int T_BdV$ D.&V$_{\text{peak}}$&FWHM&rms& $T_{kin}$&$T_{ex}$&$\tau_L$\\
    &[MHz]&MAJ[$''$]$\times$ MIN[$''$] (PA[$^{\circ}$]) &[K.km.s$^{-1}$]&[K.km.s$^{-1}$]& [km.s$^{-1}$] & [km.s$^{-1}$]& [K] & [K]&[K]&\\
     \hline
     \multicolumn{11}{c}{CSO33-b-a}\\
        \hline
       \multirow{4}{*}{CH$_3$OH} &218440& 0.52 $\times$ 0.29 (106)& 18.5 $\pm$ 2.6 &17.2 $\pm$ 2.2 & 9.4  $\pm$ 0.5 & 5.7 $\pm$ 0.7  &0.6 & \multirow{4}{*}{105}&126& 6$.10^{-2}$\\
        & 234683&0.43 $\times$ 0.41 (-27)&6.0 $\pm$ 1.2 & 5.2$\pm$ 1.2&8.2 $\pm$ 0.5&3.4 $\pm$ 1.0&0.4&&101&2.4$.10^{-3}$ \\
        &243915 &  0.32 $\times$ 0.28 (101)&26.8 $\pm$ 3.1& 25.6 $\pm$ 2.6&9.2 $\pm$ 0.5&5.1 $\pm$ 0.5 &0.4&&105&9$.10^{-2}$\\
        &\textit{261805} &\textit{0.29 $\times$ 0.25 (-78)}&\textit{19.4 $\pm$ 2.6}&\textit{ 17.3 $\pm$ 2.0 }&\textit{ 9.2 $\pm$ 0.5}&\textit{6.7 $\pm$ 0.7}&\textit{0.5}&&\textit{112} &\textit{4$.10^{-2}$}\\
        \hline
     \multicolumn{11}{c}{FIR6c-a}\\
        \hline
        \multirow{6}{*}{CH$_3$OH}&\textit{218440}& \textit{0.52 $\times$ 0.29 (107)}&\textit{10.4 $\pm$ 1.4}&\textit{11.2 $\pm$ 1.3}&\textit{11.2 $\pm$ 0.1}&\textit{2.5 $\pm$ 0.4}&\textit{0.6 }&\multirow{6}{*}{180}&\textit{13200}&\textit{5$.10^{-2}$}\\
        & 232945 & 0.48 $\times$ 0.27 (-71)&9.3 $\pm$ 1.4&8.9 $\pm$ 1.6&10.8 $\pm$ 0.2&4.4 $\pm$ 0.6&0.2&&44.8&5.2\\
        & 234683&0.47 $\times$  0.27 (109)&10.2 $\pm$ 2.2&8.6 $\pm$ 1.7 &11.4 $\pm$ 0.3&3.6 $\pm$ 0.8&0.5 &&50.3&3.9\\
        &234698&0.47 $\times$  0.27 (109)&6.7 $\pm$ 1.5&5.2 $\pm$ 0.9 &10.9 $\pm$ 0.6&3.3 $\pm$ 1.1&0.5&&39.6&1.1\\
        &\textit{243915} &\textit{0.32 $\times$ 0.27 (-78)}&\textit{22.4 $\pm$ 2.6}&\textit{20.4 $\pm$ 2.2}&\textit{10.7 $\pm$ 0.1}&\textit{3.2 $\pm$ 0.3}&\textit{0.5}&&\textit{169}&\textit{4.8}\\
        &\textit{261805} &\textit{0.30 $\times$ 0.25 (-77)}&\textit{16.6 $\pm$ 2.6}&\textit{12.1 $\pm$ 1.4}&\textit{10.8 $\pm$ 0.4 }&\textit{4.0 $\pm$ 1.0 }&\textit{0.5}&&\textit{99.5}&\textit{3.4}\\
        \hline
     \multicolumn{11}{c}{MMS9-a }\\
        \hline
       \multirow{6}{*}{CH$_3$OH} &\textit{218440}& \textit{0.53 $\times$ 0.29 (107)} &\textit{51.9 $\pm$ 5.4}&\textit{ 51.0 $\pm$ 5.3} &\textit{11.0 $\pm$ 0.5 }&\textit{ 6.7 $\pm$ 0.5  }&\textit{0.6}&\multirow{7}{*}{170}&\textit{1130}&\textit{0.9}\\
        & 232945 & 0.49 $\times$ 0.27 (-71) &40.8 $\pm$ 4.4& 39.2 $\pm$ 4.1 & 11.8 $\pm$ 1.2 & 7.5 $\pm$ 1.2 &0.3&&70.1&5.7\\
        & 234683&0.48 $\times$ 0.27 (109)&40.8 $\pm$ 4.6& 40.0 $\pm$ 4.2 & 11.1 $\pm$ 0.5 & 6.2 $\pm$ 0.5&0.5&&75&4.4\\
        &234698&0.48 $\times$ 0.27 (109)&28.1 $\pm$ 3.7& 27.3 $\pm$ 3.0 & 10.9  $\pm$ 0.5& 6.1 $\pm$ 0.6 &0.5&&73.6&1.2\\
        &\textit{243915} & \textit{0.32 $\times$ 0.27 (-256)} &\textit{89.6 $\pm$ 9.1}& \textit{85.8  $\pm$ 8.7}&\textit{ 11.1 $\pm$ 0.5} &\textit{6.7 $\pm$ 0.5}&\textit{0.5 }&&\textit{165}&\textit{7.8}\\
        &\textit{261805} &\textit{0.30 $\times$ 0.25 (-75)}&\textit{62.9 $\pm$ 6.7}& \textit{60.3 $\pm$ 6.1} &\textit{ 11.8 $\pm$ 0.5} & \textit{7.0  $\pm$ 0.5} &\textit{0.5}&&\textit{122}&\textit{4.4}\\
        \hline
     \multicolumn{11}{c}{MMS5 }\\
        \hline
       \multirow{6}{*}{CH$_3$OH} &\textit{218440}& \textit{0.52 $\times$ 0.3 (107)} & \textit{73.2 $\pm$ 7.4} &\textit{73.9 $\pm$ 7.5}&\textit{ 10.4 $\pm$  0.5}&\textit{3.2 $\pm$ 0.5} & \textit{0.6}&\multirow{9}{*}{105}&\textit{139}&\textit{11.2}\\
        & 232945 & 0.48 $\times$ 0.28 (-71)&52.5 $\pm$ 5.4 &52.2 $\pm$5.2& 10.3 $\pm$ 1.2 & 3.5 $\pm$ 1.2& 0.3&&93.5&4.2\\
        &234683&0.46 $\times$ 0.27 (-71)&  65.3 $\pm$ 6.7 &65.6 $\pm$ 6.7& 10.3 $\pm$ 0.5 & 3.2  $\pm$ 0.5&0.6&&93.5&5.0\\
        & 234698&0.46 $\times$ 0.27 (-71)&49.2  $\pm$ 5.1 &49.2 $\pm$ 5.0& 10.3 $\pm$ 0.5 & 3.0 $\pm$ 0.5& 0.6&&120&0.9\\
        &\textit{243915 }&\textit{0.32  $\times$ 0.28 (-78)}& \textit{102.8 $\pm$ 10.3}&\textit{103.2 $\pm$ 10.4}& \textit{10.3  $\pm$ 0.5}& \textit{3.4 $\pm$ 0.5}&\textit{0.5}&&\textit{105}&\textit{17.7}\\
        &\textit{261805} &\textit{0.30 $\times$ 0.25 (-77)} & \textit{96.1  $\pm$ 9.8}&\textit{96.5 $\pm$ 9.9}& \textit{10.3 $\pm$ 0.5 }&\textit{ 3.1 $\pm$ 0.5 }& \textit{1.0}&&\textit{110}&\textit{7.6}\\
        \cline{1-8}
        \cline{10-11}
        CH$_3^{18}$OH & 231758 &0.48 $\times$ 0.28 (-71)&3.5 $\pm$ 1.3&3.3 $\pm$ 1.3&10.3 $\pm$ 1.2 &2.6 $\pm$ 1.1 &0.3& &108&6.$10^{-2}$\\
         \hline
     \multicolumn{11}{c}{SIMBA-a }\\
        \hline
        \multirow{3}{*}{CH$_3$OH}&218440& 0.52 $\times$ 0.3 (106)&2.6 $\pm$ 0.8 &3.0 $\pm$ 0.7&13.0 $\pm$ 0.3 &2.0 $\pm$ 0.7 &0.5&\multirow{3}{*}{190}&195&-7.$10^{-2}$\\
        &243915 & 0.32 $\times$ 0.28 (-259)&4.7 $\pm$ 1.0&5.0 $\pm$ 1.0&13.2 $\pm$ 0.4 &2.2 $\pm$ 0.6 &0.4&&172&9.$10^{-2}$\\
        &261805 &0.30 $\times$ 0.26 (-78)&1.9 $\pm$ 1.0&1.7 $\pm$ 0.6 &13.1 $\pm$ 0.3 &1.8 $\pm$ 0.6 &0.4&&3150&-2.5.$10^{-3}$\\
        \hline
    \end{tabular}}
    \tablecomments{Results of the Gaussian fit (G.) and of the direct integration of channel intensities (D.) for the integrated intensities are reported in Cols. 4 and 5, respectively. The calibration uncertainty of 10\% has been included in the line intensity errors. $T_{kin}$ is the best fit for the kinetic temperature obtained from the LVG analysis, and $T_{ex}$ and $\tau_{L}$ are the associated excitation temperature and line optical depth. The \textit{italic} lines are those that were \textit{not} taken into account in the LTE and LVG analyses.}
\end{table*}

\section{LTE analysis: Rotational diagrams}\label{appdx:LTE}
We show here the rotational diagram (RD) obtained for each source. We can clearly see that the line at 45.4 K is masing and that points are scattered due to optically thick and/or non-LTE effects.

\begin{figure*}
\renewcommand\thefigure{C.1}
    \centering
    \includegraphics[width=0.8\linewidth]{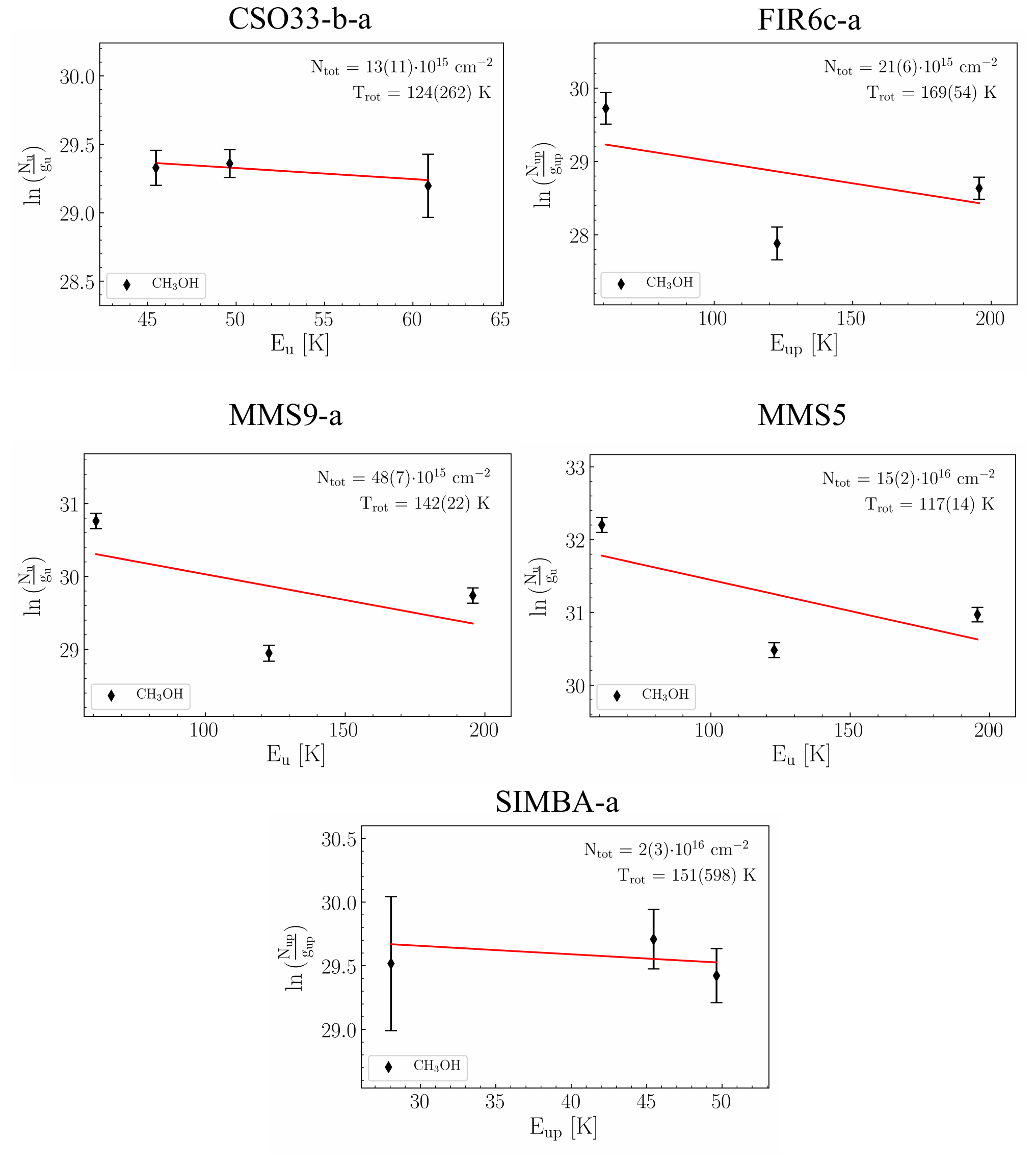}
    \caption{Rotational diagrams. Non-LTE and optically thick effects are clearly visible in FIR6c-a, MMS9-a and MMS5, as the points are scattered through the plots.}
    \label{fig:RDs}
\end{figure*}



\end{document}